\documentclass[11pt]{article}

\usepackage{LatexCustom}

\title{}
\author{}

\begin{document}

%\maketitle
    
\vspace{1truecm}
\renewcommand{\thefootnote}{\fnsymbol{footnote}}
\begin{center}

{\huge \bf{The Schwinger-Keldysh Coset Construction}}

\end{center} 

\vspace{1truecm}
\thispagestyle{empty}
\centerline{\Large Can Onur Akyuz${}^{\it a}$,\footnote{\href{mailto:cakyuz@andrew.cmu.edu}{\tt cakyuz@andrew.cmu.edu}} Garrett Goon${}^{\it a,b}$,\footnote{\href{mailto:garrett.goon@hpe.com}{\tt garrett.goon@hpe.com}} and Riccardo Penco${}^{\it a}$\footnote{\href{mailto:rpenco@cmu.edu}{\tt rpenco@andrew.cmu.edu}} }
\vspace{.7cm}

\centerline{\it ${}^{\it a}$Department of Physics, Carnegie Mellon University}
\centerline{\it 5000 Forbes Ave, Pittsburgh, PA 15213, USA}

\vspace{0.3cm}

\centerline{\it ${}^{\it b}$Determined AI, Hewlett Packard Enterprise}

\vspace{1.5cm}

\begin{abstract}

The coset construction is a tool for systematically building low energy effective actions for Nambu-Goldstone modes.  This technique is typically used  to compute time-ordered correlators appropriate for $S$-matrix computations for systems in their ground state. In this paper, we extend this technique to the Schwinger-Keldysh formalism, which enables one to calculate a wider variety of correlators and applies also to systems in a mixed state. We focus our attention on internal symmetries and demonstrate that, after identifying the appropriate symmetry breaking pattern, Schwinger-Keldysh effective actions for Nambu-Goldstone modes can be constructed using the standard rules of the coset construction.  Particular emphasis is placed on the thermal state and ensuring that correlators satisfy the KMS relation. We also discuss explicitly the power counting scheme underlying our effective actions. We comment on the similarities and differences between our approach and others that have previously appeared in the literature.  In particular, our prescription does not require the introduction of additional ``diffusive'' symmetries and retains the full non-linear structure generated by the coset construction.  We conclude with a series of explicit examples, including a computation of the finite-temperature two-point functions of conserved spin currents in non-relativistic paramagnets, antiferromagnets, and ferromagnets. Along the way, we also clarify the discrete symmetries that set antiferromagnets apart from ferromagnets, and point out that the dynamical KMS symmetry must be implemented in different ways in these two systems.

\end{abstract}

\newpage

\tableofcontents

\renewcommand*{\thefootnote}{\arabic{footnote}}
\setcounter{footnote}{0}

\newpage

\section{Introduction} \label{sec:Introduction}

The fundamental symmetries of the universe are not always easy to discern. For instance, our immediate surroundings often obscure the underlying principle that there is no preferred location.\footnote{Despite the claims of many New Yorkers.} Even when a symmetry is hidden by circumstances, {\it i.e.} it is \textit{spontaneously broken}, it still constrains the dynamics of a system and it leads to observable consequences. One robust implication of spontaneous symmetry breaking (SSB), encapsulated in Goldstone's celebrated theorem, is the existence of gapless modes which often dominate the behavior of the system at large enough length scales.  

Spontaneously broken symmetries act non-linearly on the Goldstone fields, and for this reason it is often non-trivial to identify all possible invariant operators that should be included in the low-energy effective action. Fortunately, there exists a systematic procedure to achieve this, which goes under the name of \textit{coset construction}~\cite{Coleman:1969sm,Callan:1969sn,Volkov:1973vd,Ogievetsky}.  In the simplest scenarios, this method takes as its only input the symmetry breaking pattern $G \to H$, where $G$ is the fundamental symmetry group, and $H$ is the subgroup of symmetries that remain manifest and are realized linearly. The output of this construction is a series of building blocks that can be easily combined to write down the most general local effective action $\SEFT[\pivec]$ describing the Goldstone modes $\pivec$. Zero-temperature, time-ordered correlators and the corresponding $S$-matrix elements can then be calculated using the standard path-integral representation of the generating functional:
\begin{align}\label{InOutGenerator}
Z[J]&\equiv \int\Dcal \pivec\, e^{i\SEFT[\pivec]+i\int\rd^{d+1}x\, \vec{J}(x)\cdot\pivec(x)} . 
\end{align}

Generating functionals of the form \eqref{InOutGenerator} are appropriate for so-called \textit{in-out} calculations, in which the state of the system is specified both at early and late times~\cite{Galley:2012hx}. Scattering events are the prototypical example of this scenario. There are however many other physical quantities which require instead in-in calculations, in which the (pure or mixed) state of a system is only specified at some initial time, and afterwards the system evolves according to its own dynamics.  The observables relevant for cosmology, hydrodynamics, and generic dissipative systems, for example, are most naturally phrased within this latter framework, and finite-temperature effects can be included in a natural manner.   In this setting, calculations are instead often performed using the \textit{Schwinger-Keldysh} generating functional, whose path-integral representation requires a \textit{doubling} of fields, the analogue of \eqref{InOutGenerator} being
\begin{align}
Z[J_{1},J_{2}] &\equiv \int\Dcal \pivec_{1}\Dcal \pivec_{2}\, e^{i\SEFT[\pivec_{1},\pivec_{2}]+i\int\rd^{d+1}x\, \vec{J}_{1}(x)\cdot\pivec_{1}(x)- \vec{J}_{2}(x)\cdot\pivec_{2}(x)} \ .\label{InInGeneratorSchematic}
\end{align}

In this paper, we investigate to what extent the traditional coset construction can be used to write down the doubled-field effective action $\SEFT[\pivec_{1},\pivec_{2}]$ that defines the Schwinger-Keldysh generating functional~\eqref{InInGeneratorSchematic}. In the case of Goldstone fields, doubling the field content would naively appear to be in tension with non-linear realization of the spontaneously broken symmetries. We will explicitly address this by providing a systematic prescription for writing down all possible operators that involve two copies of the Goldstone fields and are compatible with all the symmetries.  Our results provide a complementary perspective on the modern field-theoretic description of non-equilibrium systems, reviewed recently in, e.g., \cite{Glorioso:2018wxw}, as the ingredients in both constructions are intimately related.  In particular, we study the additional constraints that a finite temperature places on top of the traditional coset-construction rules, as encapsulated by the discrete \textit{dynamical Kubo-Martin-Schwinger} (DKMS) symmetry argued for in \cite{Crossley:2015evo}.  Previous studies of the coset construction within the Schwinger-Keldysh formalism can be found in~\cite{Hongo:2019qhi,Landry:2019iel}, for instance. See also~\cite{Minami:2018oxl,Hidaka:2019irz,Minami:2020ppp} for interesting discussions of Goldstone modes in out-of-equilibrium systems.

The rest of this paper is organized as follows. In order to make our discussion as self-contained as possible, we will start by reviewing the basics of the Schwinger-Keldysh formalism in Sec. \ref{sec:InInEFT}. We will focus in particular on those subtle aspects that will be relevant for the remainder of the paper. Readers that are already familiar with the Schwinger-Keldysh formalism can skip most of this section, but might still benefit from reading subsections in \ref{sec:InInEFT}. We will then turn our attention to the coset construction in Sec.~\ref{sec:InOutCoset}. We will first review the basics of this method and then, in Sec. \ref{sec:coset in-in}, show how it can be applied to Schwinger-Keldysh effective actions, focusing specifically on internal symmetries. This section introduces all the rules of the game that one should follow to write down operators that are compatible with all the symmetries. We will also discuss how turning on a finite temperature impacts the symmetries of the system, and introduces new expansion parameters in the effective theory. Finally, in Sec. \ref{sec:Examples} we illustrate our main points with a variety of explicit examples. Some supplementary material is provided in the Appendices.\\

\noindent \textit{Conventions:} Unless otherwise specified, we work in units such that $\hbar = k_B = 1$. The vast majority of our results are independent of the Minkowski metric, but wherever it matters we adopt a mostly plus signature. We define Fourier transforms as $f(\omega,  \vec k )=\int dt d^3 x\, e^{i \omega t - i \vec k\cdot \vec x} f(t, \vec x)$.  Path integrals over spatial slices versus those over all of spacetime are notationally distinguished via their respective measures:  $D \varphi \equiv \prod_{\vec x} d \varphi(t,\vec x)$ versus $\Dcal \varphi \equiv \prod_{t, \vec x} d \varphi(t, \vec x)$.  States at fixed asymptotic time slices are denoted as in $|\varphi, \pm \infty\rangle$.

%%%%%%%%%%%%%%%%%%%%%%%%%%%%%%%%%%%%%%%%%%%%%%
%%%%%%%%%%%%%%%%%%%%%%%%%%%%%%%%%%%%%%%%%%%%%%

\newpage

\section{The Schwinger-Keldysh Formalism from an EFT perspective} \label{sec:InInEFT}

In this section, we review the Schwinger-Keldysh (or in-in) formalism from an EFT perspective. We will discuss the main features of this formalism (doubling of the field content, implementation of symmetries, power counting, etc.) and define the notation that we will use throughout the paper. We will strive to keep our remarks as general as possible, so that they equally apply to systems with or without~SSB.

\subsection{Field Content} \label{sec: field content}

Consider a \emph{closed} quantum system described by some fields collectively denoted by $\varphi$ and corresponding action $S[\varphi]$, and that is in a state specified by the density matrix $\rho$.\footnote{We are working in the Heisenberg picture, where operators and their eigenstates evolve in time, while the state of the system does not.} The most general observables of the system are phrased in terms of correlation functions of the form 
\begin{align}
\langle \Ocal^{(n)}(x_{n})\ldots \Ocal^{(1)}(x_{1})\rangle&\equiv \Tr \rho \, \Ocal^{(n)}(x_{n})\ldots \Ocal^{(1)}(x_{1}) \ ,
\end{align} 
where the $\Ocal^{(i)}$'s are local operators built out of the fields $\varphi$ and their derivatives evaluated at points $x_{i}$ that are not in any particular order.   In many applications, one is interested in (linear combinations of) the restricted set of observables where the operator product is of the following factorized form:
\begin{align}
\langle \bar{T}\left [\Ocal^{(n)}(x_{n})\ldots \Ocal^{(m+1)}(x_{m+1})\right ]T\left [\Ocal^{(m)}(x_{m})\ldots\Ocal^{(1)}(x_{1})\right ]\rangle\ ,\label{FactoredCorrelator}
\end{align}
where $T$ and $\bar{T}$ denote time-ordered and anti-time-ordered products, respectively. Specializing further to the case of a single operator $\Ocal(x)$ for clarity of presentation, correlators of this form can be obtained systematically by differentiating a generating functional that depends on two external currents:
\begin{align}
Z[J_{1},J_{2}] &\equiv \Tr \left[ \rho\, \bar{T} e^{-i \int J_{2} \Ocal} \, T e^{i \int J_{1} \Ocal}   \right] . \label{ConnectedGeneratorFromTrace}
\end{align}
Generators for correlators of more than one operator or with more general time-orderings can be similarly constructed, but the one in Eq. \eqref{ConnectedGeneratorFromTrace} will be sufficient for our purposes.

Inserting resolutions of the identity ${\bf 1}=\int D\varphi\, \ket{\varphi, \pm \infty}\bra{\varphi,\pm \infty}$ between the various factors inside the trace, we can rewrite the generating functional as
\begin{align}
Z[J_{1},J_{2}] &= \! \int \! D\varphi_{a} D \varphi_{b} D \varphi_{c}\,\bra{\varphi_{a}, -\infty} \rho\ket{\varphi_{c}, -\infty}\bra{\varphi_{c}, -\infty}\bar{T} e^{-i \int J_{2} \Ocal}\ket{\varphi_{b}, +\infty} \label{ConnectedGeneratorPIInsertions} \\ 
& \qquad \qquad \qquad \qquad \qquad \qquad \qquad \qquad \qquad \times \bra{\varphi_{b}, +\infty}T e^{i \int J_{1} \Ocal}\ket{\varphi_{a}, -\infty} , \nonumber
\end{align}
where the last factor admits a path integral representation of the form
\begin{align}
\bra{\varphi_{b}, + \infty}  T e^{i \int J_{1} \Ocal}\ket{\varphi_{a}, -\infty} &=\int_{\varphi(-\infty)=\varphi_{a}}^{\varphi(+\infty)=\varphi_{b}}\Dcal\varphi\, e^{iS[\varphi]+i\int\rd^{d+1}x\, J_{1}\Ocal}\ , \label{AmpltudePathIntegral}
\end{align}
and a similar, conjugated expression holds for the second factor.

It is customary for the preceding construction to be summarized in the compact form
\begin{align}
Z[J_{1},J_{2}] &= \int_\rho \Dcal\varphi_{1}\Dcal\varphi_{2}\, \exp\left[iS[\varphi_{1}]-iS[\varphi_{2}]+i\int\rd^{d+1}x\, J_{1}\Ocal (\varphi_1) -J_{2}\Ocal(\varphi_2)\right] \ .\label{ConnectedGeneratorPICompact}
\end{align}
We see therefore that calculating correlation functions of the form \eqref{FactoredCorrelator} requires a doubling of the field content: this is a direct consequence of the fact that expectation values, as opposed to transition amplitudes, require time-evolving \textit{both} the corresponding bras and kets, with one set of fields performing each such evolution---see~\cite{Galley:2012hx} for a related discussion.

\subsection{Effective Action} \label{sec:effective action}

The expression in Eq. \eqref{ConnectedGeneratorPICompact} leaves the dependence on the density matrix $\rho$ and boundary conditions at $t= +\infty$ highly implicit. These are nonetheless important features of the path integral formulation that will play an important role in what follows. 

The generator \eqref{ConnectedGeneratorPICompact} corresponds to a path integral over a contour $\Ccal$ which extends from $t=-\infty$ to $t=+\infty$ and back, with appropriate boundary conditions placed in the asymptotic regions; see for instance Fig.~\ref{fig:InInTimeContourFiniteBeta} in appendix \ref{app:FreeInIn} for a depiction of this contour in the case where $\rho$ is thermal.  Consequently, $Z[J_{1},J_{2}]$ generates correlators which are \textit{path-ordered} along $\Ccal$, for the usual reasons, with operators labeled with a 2 always coming after those labeled with a 1.  For example, when $\mathcal{O}(\varphi) = \varphi$, the various permutations of the two-point function are explicitly given by
 \begin{align}
 \langle \Pcal \varphi_{i}(x)\varphi_{j}(x')\rangle=\begin{cases}
 \langle T\varphi(x)\varphi(x')\rangle & \quad i=j=1\\
  \langle \bar{T}\varphi(x)\varphi(x')\rangle & \quad i=j=2\\
   \langle \varphi(x)\varphi(x')\rangle & \quad i=2,j=1\\
    \langle \varphi(x')\varphi(x)\rangle & \quad i=1,j=2
 \end{cases}\ ,\label{TwoPointFunctions}
 \end{align}
 where $\Pcal$ denotes path-ordering.  For brevity, we will omit explicit $\Pcal$'s in subsequent expressions and path-ordering is always implied unless noted otherwise.

 The shorthand expression \eqref{ConnectedGeneratorPICompact} is cryptic, at best.  A particular shortcoming is that the correlator \eqref{TwoPointFunctions} does not arise from simply inverting the kinetic term of $S[\varphi_{1}]-S[\varphi_{2}]$, as the notation might suggest.  Were this true, all mixed correlators such as $\langle \varphi_1(x)\varphi_2(x')\rangle$ would necessarily be vanishing.  Instead, the boundary conditions implicit in \eqref{ConnectedGeneratorPICompact} generate the necessary, non-trivial cross-couplings---see App.~\ref{app:FreeInIn} for an explicit example.
 
 Nuisances such as this one motivate the use of an alternative description based on an \emph{effective action} $\SEFT[\varphi_1,\varphi_2]$ in which the boundary conditions implicit in \eqref{ConnectedGeneratorPICompact} are made explicit and encoded into $\SEFT$ itself. These effective actions are of the functional form
\begin{align} \label{effective action closed}
	\SEFT[\varphi_{1},\varphi_{2}] = S[\varphi_{1}]-S[\varphi_{2}] + \Delta S[\varphi_1, \varphi_2] \ , 
\end{align}
where the information about the state $\rho$ of the closed system as well as the boundary conditions are encoded in the cross terms $\Delta S[\varphi_1, \varphi_2]$~\cite{KamenevBook2011,Glorioso:2018wxw}.\footnote{Note that the fields $\varphi$ that appear in the effective action \eqref{effective action closed} generically do not coincide with the fields $\varphi$ in Eq. \eqref{ConnectedGeneratorPICompact}---a standard observation in more traditional in-out effective field theories. Similarly, the functional $S[\varphi_{i}]$ in Eq. \eqref{effective action closed} will generally differ from the functional $S[\varphi_{i}]$ in Eq. \eqref{ConnectedGeneratorPICompact}; we use the same symbol in both contexts to avoid unnecessary additional notation.} We elaborate upon this point later in this section and will also discuss the circumstances under which we expect $\SEFT$ to be local.  Correlators are computed using $\SEFT$ as in
\begin{align}
    \langle \Ocal^{(n)}(x_n)\ldots\Ocal^{(1)}(x_1) \rangle = \int\Dcal\varphi_1\Dcal\varphi_2\, e^{i\SEFT[\varphi_1, \varphi_2]}\Ocal^{(n)}(x_n)\ldots\Ocal^{(1)}(x_1)\ , 
\end{align}
where perturbative computations now proceed in the naive manner in which free propagators are determined by the quadratic terms in $\SEFT$ and non-linearities are handled order by order.

A term of the form $\Delta S[\varphi_1, \varphi_2]$ arises also in \emph{open} systems, albeit in a different way~\cite{Feynman:1963fq,Calzetta:2008iqa,Sieberer:2015svu,Haehl:2016pec}. To see this, imagine decomposing our closed system in a subsystem of interest, with degrees of freedom $\phi$, and an environment, with degrees of freedom $\Phi$. If we are only interested in correlation functions of the subsystem, we can work with a generating functional of the form
\begin{align}
Z[J_{1},J_{2}] &= \int_\rho \Dcal\phi_{1}\Dcal\phi_{2}\Dcal\Phi_{1}\Dcal\Phi_{2}\, e^{iS[\phi_{1},\Phi_{1}]-i S[\phi_{2},\Phi_{2}]+i\int\rd^{d+1}x\, J_{1}\Ocal (\phi_1) -J_{2}\Ocal(\phi_2)}\ ,
\end{align}
where $S [\phi, \Phi] = S [\phi] + S_{\rm int} [\phi, \Phi]$, and $S_{\rm int} [\phi, \Phi]$ captures the dynamics of the environment as well as its interactions with the subsystem. Assuming for simplicity that the state of the full system is factorized, i.e. $\rho = \rho_\phi \otimes \rho_\Phi$, we can integrate out the fields $\Phi$ to obtain a correction to the action for the subsystem given by
\begin{align} \label{influence functional}
	e^{i \Delta S[ \phi_1, \phi_2]} = \int_{\rho_\Phi} \Dcal\Phi_{1}\Dcal\Phi_{2}\,e^{iS_{\rm int}[\phi_{1},\Phi_{1}]-i S_{\rm int}[\phi_{2},\Phi_{2}]} . 
\end{align}
The term $\Delta S[ \phi_1, \phi_2]$ that arises by integrating out the environment is known as the \emph{influence functional}, and it generically includes interactions between $\phi_1$ and~$\phi_2$. The origin of these interactions can be traced back to the ``off-diagonal'' correlation functions of the environment degrees of freedom, e.g. $\langle \Phi_i \Phi_j \rangle$ with $i \neq j$. Alternatively, one could also derive the influence functional using the fact that, for open systems, the evolution between subsequent time slices is not unitary and is performed by the Lindblad operator---see e.g.~\cite{Sieberer:2015svu}.

Eq. \eqref{influence functional} shows explicitly that, in the case of an open system, the term  $\Delta S[ \phi_1, \phi_2]$ contains information about the state of the environment $\rho_\Phi$ and its interactions with the subsystem. This is conceptually different from the effective action of a closed system in \eqref{effective action closed}, which instead depends on the state of the system under consideration. Of course, one could amend the influence functional so that it also captures the state $\rho_\phi$ of the subsystem, following the same strategy we outlined for a closed system. By doing so, $\Delta S[\phi_1, \phi_2]$ would now encode once again the state of a closed system, $\rho = \rho_\phi \otimes \rho_\Phi$. This procedure would be equivalent to first introducing the effective action for the total closed system, and then integrating out the environment in the naive manner:
\begin{align}
	\int_\rho \Dcal \phi_1 \Dcal \phi_2 \Dcal \Phi_1 \Dcal \Phi_2 e^{iS[\phi_{1},\Phi_{1}]-i S[\phi_{2},\Phi_{2}]} &= \int \Dcal \phi_1 \Dcal \phi_2 \Dcal \Phi_1 \Dcal \Phi_2 e^{iS[\phi_{1},\Phi_{1}]-i S[\phi_{2},\Phi_{2}] + i \Delta S[\phi_1, \phi_2,  \Phi_1, \Phi_2 ]} \nonumber \\
	&= \int \Dcal \phi_1 \Dcal \phi_2 \, e^{iS[\phi_{1}]-i S[\phi_{2}] + i \Delta S[\phi_1, \phi_2]} .
\end{align}
From this perspective, the difference between an open and a closed system is whether the degrees of freedom appearing in the path integral define a basis for the Hilbert space on which the state $\rho$ is defined.

\subsection{Locality and Expansion Parameters}  \label{sec:locality}

Let's consider an EFT described by an ``in-out'' effective action $\SEFT[\varphi]$ with energy cutoff $\Lambda$.\footnote{In non-relativistic EFTs one should distinguish between the cutoffs for energy and momentum---as we will see in the examples discussed in Sec. \ref{sec:Examples}. In this section we focus our attention on the energy cutoff to simplify the discussion. } This means that any operator in the in-out effective action can be assigned a definite scaling in the ratio $E /\Lambda$, with $E$ the typical scale of interest~\cite{Rothstein:2003mp,Penco:2020kvy}. Any state $\rho$ that is not the vacuum of the EFT will generically introduce additional scales into the problem. We will collectively denote these scales with $M$; some examples are: the temperature $T$ of a thermal state, the chemical potential $\mu$ of a finite density state, the characteristic length scale $\ell$ of a semi-classical field profile (converted to an energy scale using some characteristic speed), etc. Only states such that $M \ll \Lambda$ can be reliably described within the in-out EFT.

In the approach described in Sec. \ref{sec:effective action}, these additional scales $M$ appear explicitly in the Schwinger-Keldysh effective action $\SEFT[\varphi_1, \varphi_2]$. While it is perhaps plausible that such an action might be able to reproduce all correlators of the form \eqref{FactoredCorrelator}, there is certainly no expectation that this EFT should be local for all scales $E \ll \Lambda$---and indeed, it generically won't be. In order to work with a local Schwinger-Keldysh effective action, we need to restrict ourselves to the regime $E\ll M \ll \Lambda$, which will be the focus of this paper. Below the scales $M$, all information about the state $\rho$ is encoded in the effective action by an infinite tower of irrelevant local operators---the usual way in which UV physics manifests itself at low energies. Thus, the Schwinger-Keldysh effective action now contains a new expansion parameter, $E / M$.

This point is potentially confusing for thermal states: because EFTs at zero temperature are usually described using a local action, one might expect that the $T \ll E$ regime should also admit a local description by continuity. However, thermality is generically encoded by $\sim e^{- E/T}$ factors \cite{Niemi:1983ea,Niemi:1983nf,Landsman:1986uw} which only admit an expansion in powers of $E$ for $T\gg E$, the regime in which being at finite temperature appears as a UV effect.\footnote{We are being schematic here. More precisely, the scale $M$ is proportional to $T$ but does not necessarily coincide with it. For weakly coupled UV completions, $M$ is more accurately of order $g^4 T \ln g^{-2}$, a scale associated with large-angle scattering events. It is below this scale that the relevant degrees of freedom become the hydrodynamic modes \cite{arnold1998effective}. This situation is a manifestation of the usual fact that, for weakly coupled UV completions, the cutoff of the low-energy EFT can be parametrically smaller than the strong coupling scale (in our case, $T$).}

At scales $E \ll M$, the relevant degrees of freedom are often the Goldstone modes associated with symmetries that are spontaneously broken by the state $\rho$. In the following sections, we will uncover the principles that one should follow to write down the most general Schwinger-Keldysh effective action for such Goldstone modes.

Irrespective of the state $\rho$ of the system, Schwinger-Keldysh effective actions are also naturally endowed with another expansion parameter, namely $\hbar$, that controls the semi-classical expansion. In the absence of Goldstone modes, fields usually transform linearly under all the symmetries, and a systematic $\hbar$ expansion of the effective action is straightforward to implement~\cite{Glorioso:2018wxw} (see also Sec. \ref{sec:KeldyshRotation} for more details). In the presence of Goldstone bosons, however, an expansion of the effective action in powers of $\hbar$ requires some extra care, since a naive implementation would break some of the symmetries that are realized non-linearly, as will become clear in subsequent sections. For this reason, we will not be implementing this expansion in what follows.

\subsection{Symmetries} \label{sec:symmetries}

EFTs are specified not only by their field content and expansion parameter(s), but also by their symmetries. It is therefore important to discuss which symmetries one should impose when writing down the most general Schwinger-Keldysh effective action $\SEFT[\varphi_1,\varphi_2]$. The symmetry considerations that must inform the construction of the effective action for closed systems are:

\begin{itemize}
    \item \emph{Gauge symmetries}: When the single field action $S[\varphi]$ enjoys a \textit{gauge} symmetry, the action $S[\varphi_{1},\varphi_{2}] = S[\varphi_{1}] - S[\varphi_{2}]$ in Eq.~\eqref{ConnectedGeneratorPICompact} is invariant under independent gauge transformations of the $\varphi_{i}$'s that coincide at $t = \pm \infty$~\cite{Glorioso:2018wxw}. As a result, the Schwinger-Keldysh effective action $S_{\rm EFT}[\varphi_1, \varphi_2]$ must be invariant under two copies of the gauge group. This is consistent with the statement in the previous section that one should double all the degrees of freedom---including gauge fields. 
	\item \emph{Continuous global symmetries}: The fact that the two gauge transformations must coincide at $t = \pm \infty$ implies that, in the global limit, the action $S[\varphi_{1},\varphi_{2}] = S[\varphi_{1}] - S[\varphi_{2}]$ can only be invariant under the diagonal symmetry group $G$ which transforms the $\varphi_i$'s simultaneously. This fact is not manifest at the level of the action, since $\varphi_1$ and $\varphi_2$ would appear to be decoupled; it follows instead from the boundary conditions, and in this sense it can be viewed as a non-local constraint on the symmetries of the system. These boundary conditions, together with the state of the system, are built directly into the Schwinger-Keldysh effective action $S_{\rm EFT}[\varphi_1, \varphi_2]$. Furthermore, gauge symmetries must become physically indistinguishable from global symmetries in the limit of vanishingly small gauge coupling.\footnote{See e.g. discussion in Sec. 21.4 of~\cite{Weinberg:1996kr}.} In the regime where the Schwinger-Keldysh effective action is local, this can only be achieved if $S_{\rm EFT}[\varphi_1, \varphi_2]$ is invariant under two copies of all global symmetries, $G_1 \times G_2$.\footnote{In the ground state, $G_1 \times G_2$ is explicitly broken down to its diagonal subgroup by terms of $\Ocal(\varepsilon)$ when the ``$i \varepsilon$'' prescription is implemented at the level of the action~\cite{KamenevBook2011}. This is another manifestation of the fact that the state at $t = -\infty$ is only invariant under the diagonal symmetry group. The factors of $i \varepsilon$ are crucial to reproduce the correct $n$-point functions, but should not be taken into account when discussing the symmetries of the Schwinger-Keldysh effective action. This is standard EFT practice: for example, the in-out effective action for a $U(1)$ Goldstone is considered to be shift-invariant even though implementing the ``$i \varepsilon$'' prescription in the action would break the shift symmetry explicitly.} This enhancement of global symmetries is a direct consequence of the requirement that the Schwinger-Keldysh effective action be local and leads to the existence of two separate Ward identities, which are required to reproduce the same information of a single Ward identity defined on the two segments of the Schwinger-Keldysh contour (see Fig. \ref{fig:InInTimeContourFiniteBeta} in App \ref{app:FreeInIn}).  The symmetry properties of the matrix elements of $\rho$ in the past infinity determine to what extent $G_1 \times G_2$ is spontaneously broken down to a subgroup---we will discuss this more in depth in Sec.~\ref{sec:coset in-in}.
	\item \emph{Discrete symmetries:} The difference between past and future boundary conditions breaks explicitly time reversal, which therefore is not a symmetry of the effective action $S_{\rm EFT}[\varphi_1,\varphi_2]$ even when it is a symmetry of the single-field action $S[\varphi]$. All other discrete symmetries of the single-field action are realized twice in the Schwinger-Keldysh effective action, as is the case for continuous global symmetries. This is required for consistency, since discrete subgroups of continuous symmetries are always realized twice.
    \item \emph{Emergent symmetries:} The state $\rho$ not only determines whether some of the symmetries are spontaneously broken, but can also give rise to additional ``emergent'' symmetries in the EFT. For instance, the homogeneity and isotropy of a state would be encoded by an emergent internal $ISO(d)$ symmetry~\cite{Nicolis:2013lma,Nicolis:2015sra}. A more generic example is provided by the thermal state, which endows the effective theory with an additional discrete symmetry---the \emph{dynamical KMS symmetry} (DKMS)~\cite{Sieberer:2015hba,Glorioso:2017fpd,Glorioso:2018wxw}---that, in the simplest case,\footnote{The existence of the symmetry \eqref{KMS transformation} relies on the invariance of the underlying {\it dynamics} under time reversal. More in general, the DKMS symmetry can be implemented by combining the transformations \eqref{KMS transformation} with additional discrete symmetries, e.g. parity and charge conjugation~\cite{Glorioso:2018wxw}. The important point is that the {\it ground state} of the system must be invariant under the combined action of these discrete symmetries and time reversal. As we will see in Sec. \ref{sec:Ferromagnets}, this requirement plays an important role for ferromagnets. \label{footnote: general DKMS}} can be implemented on the two copies of the fields as follows: 

\begin{subequations} \label{KMS transformation}
\begin{align}
    \varphi_1^{\, \prime} (t, \vec x) &= \varphi_1 (-t+i \beta/2, \vec x) = \varphi_1 (-t, \vec x) - \frac{i \beta}{2} \partial_t \varphi_1 (-t, \vec x) + \Ocal(\beta^2) ,\\
    \varphi_2^{\, \prime} (t, \vec x) &= \varphi_2 (-t- i \beta/2, \vec x) = \varphi_2 (-t, \vec x) + \frac{i \beta}{2} \partial_t \varphi_2 (-t, \vec x) + \Ocal(\beta^2) ,
\end{align}
\end{subequations}
where $\beta = 1/T$ (In this paper we are aiming for an effective action up to leading order in an expansion in $E/T$, which is why we expanded the DKMS symmetry in powers of $\partial_t/T$).\footnote{We are assuming here that all continuous global symmetries act linearly on the $\varphi_i$ fields, i.e. that there is no spontaneous symmetry breaking. Furthermore, we have made an additional imaginary time-translation to bring \eqref{KMS transformation} to a convenient form---see e.g. \cite{Glorioso:2018wxw} for a more general form of these transformation rules.}  This symmetry ensures that correlation functions of the system satisfy the KMS condition~\cite{Kubo:1957mj,Martin:1959jp}, which e.g. for the 2-point correlator of any operator $\mathcal{O}$ reads 
\begin{align}
	\langle \mathcal{O}(t, \vec x) \mathcal{O}(t', \vec{x}^{\, \prime}) \rangle = 	\langle  \mathcal{O}(t'- i \beta/2, \vec{x}^{\, \prime}) \mathcal{O}(t + i \beta/2, \vec x) \rangle .
\end{align}
Furthermore,  when combined with the unitarity conditions discussed below in Sec. \ref{sec:InInUnitarity}, the DKMS symmetry leads to the existence of an entropy current with non-negative divergence~\cite{Glorioso:2016gsa}. It is worth mentioning that Eq. \eqref{KMS transformation} is not the only possible way of implementing the KMS condition as a symmetry of the effective action---see for instance~\cite{Haehl:2014zda,Haehl:2015pja}, and~\cite{Haehl:2017zac} for a detailed comparison with the approach put forward in~\cite{Glorioso:2018wxw} and adopted in this paper.
\end{itemize}

\subsection{Unitarity} \label{sec:InInUnitarity}

In the standard in-out path integral, unitarity implies that the single field effective action $\SEFT[\varphi]$ must be real. Similarly, unitarity also restricts the form of the Schwinger-Keldysh effective action $\SEFT[\varphi_1, \varphi_2]$, and requires that the following conditions be satisfied~\cite{Glorioso:2018wxw}:
\begin{enumerate}
 \item The action must vanish when $\varphi_1 = \varphi_2$: $\SEFT[\varphi,\varphi]=0$.
 \item Unlike in the usual in-out path integral, the effective action that appears in the Schwinger-Keldysh generating functional is allowed to be complex. However, under conjugation we must have $\SEFT[\varphi_{1},\varphi_{2}]^{*}=-\SEFT[\varphi_{2},\varphi_{1}]$.
 \item Furthermore, the imaginary part of the action must be non-negative: $\im \SEFT[\varphi_{1},\varphi_{2}]\ge 0$.
\end{enumerate}
In terms of the decomposition \eqref{effective action closed} $\SEFT[\varphi_{1},\varphi_{2}] = S[\varphi_{1}]-S[\varphi_{2}] + \Delta S[\varphi_1, \varphi_2]$, the first unitarity constraint implies that $\Delta S[\varphi,\varphi]=0$; the second condition requires $S[\varphi]$ to be purely real and the mixing term to obey $\Delta S[\varphi_{1},\varphi_{2}]^{*}=- \Delta S[\varphi_{2},\varphi_{1}]$; and the third unitarity condition further imposes $\im \Delta S[\varphi_{1},\varphi_{2}]\ge 0$.

\subsection{A Convenient Field Redefinition: Keldysh Rotation} \label{sec:KeldyshRotation}

In order to simplify calculations and make the causality properties of various quantities manifest, it is convenient to perform a field redefinition and switch to the so-called \emph{Keldysh basis}. Assuming for now that all symmetries are linearly realized on our fields, this is done by performing a \textit{Keldysh rotation} to introduce the following new degrees of freedom:\footnote{See \cite{Jensen:2017kzi} for a detailed discussion of subtleties associated with the Keldysh basis.} 
\begin{gather}
 \varphi_{a}(x)\equiv \varphi_{1}(x)-\varphi_{2}(x)\ , \qquad\quad \varphi_{r}(x)\equiv \tfrac{1}{2}\left [\varphi_{1}(x)+\varphi_{2}(x)\right] . \label{KeldyshBasisFields}
 \end{gather}
The Keldysh basis has several advantages. First, even though the 2-point functions of $\varphi_1$ and $\varphi_2$ are all generically non-zero, as shown in Eq. \eqref{TwoPointFunctions}, they are actually not all independent of each other. Writing out the time-orderings, one can easily show that the following identity holds~\cite{KamenevBook2011}:
 \begin{align}
 \langle \varphi_{1}(x)\varphi_{1}(x')\rangle+\langle \varphi_{2}(x)\varphi_{2}(x')\rangle-\langle \varphi_{1}(x)\varphi_{2}(x')\rangle-\langle \varphi_{2}(x)\varphi_{1}(x')\rangle = 0 \ .\label{TwoPointIdentity}
 \end{align}
In the Keldysh basis, this redundancy is made explicit through the fact that $\langle \varphi_{a} (x) \varphi_{a} (x')\rangle$ vanishes:\footnote{This pattern generalizes quite widely: any correlator involving a string of $a$- and $r$-operators $\sim\langle \varphi_{r}^{m}\varphi_{a}^{n}\rangle$ vanishes if among all operators, the one with the largest temporal argument is an $a$-operator.  This non-perturbative result is known as the \textit{largest-time equation} \cite{Veltman:1963th} and it implies, in particular, that $\langle \varphi_{a}^{n}\rangle$ is vanishing for all $n$.}\footnote{Our notation for various correlation functions and some of their properties are summarized in App.~\ref{app:CorrelatorCheatSheet}.}
 \begin{align}
 \langle \varphi_{i}(x)\varphi_{j}(x')\rangle=\begin{cases}
 \tfrac{1}{2} \langle \left\{ \varphi(x), \varphi(x') \right\}\rangle & \quad i=j=r\\
  0 & \quad i=j=a\\
   \theta (t - t') \langle [ \varphi(x), \varphi(x')]\rangle & \quad i=r,j=a\\
    \theta (t' - t) \langle [ \varphi(x'), \varphi(x) ]\rangle & \quad i=a,j=r
 \end{cases}\ .\label{TwoPointFunctionsar}
 \end{align}
  Another technical advantage to the Keldysh basis also becomes apparent in Eq. \eqref{TwoPointFunctionsar}: mixed $a$-$r$ two-point correlators have manifest causal properties.  The retarded Green's function is proportional to $\langle \varphi_{r}(x)\varphi_{a}(x')\rangle$, while the advanced one is proportional to $\langle \varphi_{a}(x)\varphi_{r}(x')\rangle$.  One can think of the mixed correlators as containing information about the system's fundamental dynamics, while $r$-$r$ correlators encode information about the state of the system.\footnote{For instance, $\langle\varphi_{r}(\omega,\vec{k})\varphi_{r}(-\omega,-\vec{k})\rangle\propto \frac{1}{2}+n_{\rm BE}(\beta\omega)$ in a thermal state, where $n_{\rm BE}(\beta\omega)$ is the Bose-Einstein distribution (assuming $\Ocal$ is bosonic)---see App.~\ref{app:FreeInIn} for more details.}

The effective action $\SEFT[\varphi_{a},\varphi_{r}]$ is generally organized as an expansion in $\varphi_{a}$ fields which, due to the form of the action in \eqref{effective action closed}, can be written as
\begin{align}
\SEFT[\varphi_{a},\varphi_{r}]&=\sum_{n\in\{1,3,5,\ldots\}}^{\infty}\left(\prod_{i=1}^{n}\int\rd^{d+1}x_{i}\right) \frac{\delta^{n}S[\varphi_{r}]}{\delta\varphi_{r}(x_{1})\ldots \delta\varphi_{r}(x_{n})}\frac{1}{2^{n-1} n!}\varphi_{a}(x_{1})\ldots \varphi_{a}(x_{n})+ \Delta S[\varphi_{a},\varphi_{r}]\ ,\label{GenericSEFTFunctionalFormKeldyshBasisFields}
\end{align}
where only odd powers of $\varphi_{a}$ can arise from the expansion of $S[\varphi_1] - S[\varphi_2]$.   In these variables, the unitarity constraints previously discussed in Sec. \ref{sec:InInUnitarity} now imply the conditions
\begin{enumerate}
	\item $\Delta S [\varphi_{a}=0,\varphi_{r}]=0$,
	\item $\Delta S[\varphi_{a},\varphi_{r}]^{*}= - \Delta S[-\varphi_{a},\varphi_{r}]$,
	\item $\im \Delta S [\varphi_{a},\varphi_{r}]\ge 0$.
\end{enumerate}
The first condition further implies that $\varphi_{a}=0$ is always a consistent solution of the equations of motion that follow from varying the effective action with respect to $\varphi_r$. Variation with respect to $\varphi_a$ yields instead, at lowest order in $\varphi_a$, the classical equations of motion for $\varphi_r$. As a result, the $a$- and $r$-fields have also the conceptual advantage of admitting a natural physical interpretation. The $\varphi_r$ operator can be identified with a classical field in the $\hbar \to 0$ limit, whereas $\varphi_a$ is responsible for quantum and stochastic effects~\cite{KamenevBook2011}. In the absence of spontaneous symmetry breaking, this can be made explicit by rescaling $\varphi_a \to \hbar \varphi_a$, so that the effective action admits a straightforward expansion in powers of $\hbar$. This procedure is consistent with semi-classical expansion of in-out effective actions in terms of some dimensionless combination of coupling constants and $\hbar$~\cite{Brodsky:2010zk} in the limit of high occupation number~\cite{Radovskaya:2020lns}. In the presence of spontaneous symmetry breaking, however, an expansion in powers of $a-$fields becomes more subtle because it obscures (i.e. break explicitly) some of the symmetries realized non-linearly. Understanding  how this relates to the semi-classical expansion of in-out effective theories with Goldstones~\cite{Panico:2015jxa} is an open problem that we leave for future study. In the present paper we will not truncate our results at a finite order in $a-$fields.

For future use, we also report here the action of the DKMS symmetry on the fields in the Keldysh basis. Working at lowest order in $E/T$, we immediately find from the transformation rules \eqref{KMS transformation}:
\begin{subequations} \label{DKMS transformations}
\begin{align}
	 \varphi_r^{\, \prime} (t, \vec x) &\simeq \varphi_r (-t, \vec x) - \tfrac{i}{4} \beta \partial_t \varphi_a (-t,\vec x), \label{phi_r KMS transformation} \\
    \varphi_a^{\, \prime} (t, \vec x) &\simeq \varphi_a (-t, \vec x) - i \beta \partial_t \varphi_r (-t, \vec x) .
\end{align}
\end{subequations}
Note that, for linearly realized symmetries, the change in $\varphi_r$ in Eq. \eqref{phi_r KMS transformation} is of $\mathcal{O}(\hbar)$, and is therefore often neglected---see e.g.~\cite{Glorioso:2018wxw}.\footnote{Note that, when restoring all factors of $\hbar$, one should replace $\beta \to \hbar \beta$ in order for the temperature to have units of energy.} We are keeping this correction here since for our purposes it will be important to distinguish between expansions in powers of $E/T$ and $\hbar$.

In the body of the paper, we focus on constructing low-energy Schwinger-Keldysh effective actions for systems in which one or more symmetries are spontanµeously broken. As we will see, working in terms of $r$- and $a$-fields becomes especially natural under these circumstances. This is because the Goldstone fields in the Keldysh basis have non-linear transformation properties that resemble those in a more traditional in-out effective theory. This will allow us, in particular, to implement the DKMS symmetry on the Goldstone fields in a way that is consistent with all the non-linearly realized symmetries. Because of the relations \eqref{KeldyshBasisFields}, instead, the Goldstone fields in the $(1,2)$ basis would satisfy much more complicated transformation rules.

\section{The Coset Construction} \label{sec:InOutCoset}

One of the tenets of modern EFTs is the principle that all the terms compatible with the symmetries of a physical system should appear in its corresponding effective action. In the presence of spontaneous symmetry breaking the identification of all such operators can be technically challenging, as the symmetries are non-linearly realized on the Goldstone modes of the system, and the standard framework for organizing the classification is the \emph{coset construction}~\cite{Callan:1969sn,Volkov:1973vd,Ogievetsky}.  Essential elements of the technique are reviewed below in the particular case of spontaneously broken \emph{internal} symmetries; a more detailed discussion can be found for instance in~\cite{Penco:2020kvy}.

\subsection{The Maurer-Cartan Form}

The first step is to distinguish those symmetries that are spontaneously broken from those that are not. Letting the breaking pattern be $G\to H$  and denoting the generators of spontaneously broken symmetries by $X_{i}$, the coset construction analysis starts by forming the \emph{coset parametrization} $\Omega\in G/H$ which can be canonically written as
\begin{align}
	\Omega = e^{i \pi^{j} (x) X_{j}}\ .\label{ReprentativeCosetElement}
\end{align}
Every broken generator\footnote{The broken generators are of course only specified up to the addition of unbroken generators---and in fact this ambiguity can often be leveraged to simplify explicit calculations. One should include in the coset parametrization only those broken generators that are not equivalent up to an unbroken transformation.} appears with an associated Goldstone field $\pi^{i}(x)$. The transformation properties of the Goldstone fields under the action of a generic symmetry transformation $g$ are defined by the relation
\begin{equation} \label{CosetTransformation}
	g \, \Omega ( \pivec) = \Omega ( \pivec^{\, \prime})\, h (g, \pivec),
\end{equation}
where $h$ is some element of the unbroken subgroup which could in principle depend on $g$ and the fields $\pi^i (x)$.

The transformation laws for $\pi^{i}$ encapsulated by \eqref{CosetTransformation} are in general non-linear and contain infinitely many terms when expanding in fields.  Therefore, the operators which respect the requisite symmetries are severely constrained.  The central building block for constructing such operators is the \emph{Maurer-Cartan 1-form}, $\bfomega =  \Omega^{-1} {\rm d} \Omega$, which is a Lie-algebra valued 1-form that can be conveniently written as
\begin{align} \label{MCForm}
	\bfomega =  \Omega^{-1} \partial_\mu \Omega \, {\rm d}  x^\mu  \equiv i\, \left(  D_{\mu}\pi^{i} X_i + \Acal_\mu^B T_B \right) {\rm d}  x^\mu  \ ,
\end{align}
where we have denoted unbroken generators with $T_B$.

The Maurer-Cartan form is useful because the above coefficients enjoy relatively simple transformation properties under the relevant $G$-symmetries \cite{Ogievetsky}:
\begin{itemize}
\item The $D_{\mu} \pi^{i}$ factors are non-linear combinations of the Goldstone fields of the form $D_{\mu} \pi^{i} = \partial_{\mu} \pi^{i} + \Ocal(\pi^2)$ which transform linearly under \eqref{CosetTransformation}:
\begin{align} 
	D_{\mu} \pi^{i} \quad \longrightarrow \quad h(g, \pivec)^i{}_j D_{\mu} \pi^j \ , \label{GoldstoneCovDTransformation}
\end{align}
where $h(g, \pivec)^i{}_j$ is a (possibly reducible) representation of the unbroken transformation $h(g, \pivec)$.  Because of $D_{\mu} \pi^{i}$'s linear transformation property \eqref{GoldstoneCovDTransformation}, it is  conventional to refer to this building block as a ``covariant derivative", but we wish to emphasize that this terminology is somewhat misleading: because of $\pi^{i}(x)$'s highly non-linear transformation laws \eqref{CosetTransformation} one cannot interpret $D_{\mu} \pi^{i} (x)$ as a combination of a partial derivative and a connection acting on $\pivec(x)$ in any standard manner.
\item The $\Acal_\mu^B$ coefficients transform like connections under $h(g, \pivec)$ and can be used to define the following covariant derivative:
\begin{align}
	\covD_{\mu} = (\partial_\mu + i \Acal_\mu^B T_B)\ .\label{GoldstoneMatterCovD}
\end{align}
We use the $\covD_\mu$ to symbol to represent proper covariant derivatives, which distinguish them from the $D_\mu$ notation used for Goldstone ``covariant derivative" defined above.  This derivative can act on any operator $\Ocal^I$ in a linear representation of $h(g, \pivec)$---be that a Goldstone covariant derivative, some matter field, or a combination thereof---and it yields a quantity that once again transforms linearly: 
\begin{align}
	\covD_{\mu} \Ocal^I \quad \longrightarrow \quad h(g, \pivec)^J{}_I \covD_{\mu} \Ocal^I . 
\end{align}
\end{itemize}
The most general $G$-symmetric in-out effective action describing Goldstone modes and their interactions with matter fields $\Psi^I$ is then of the functional form
\begin{align}
	S = \int {\rm d}^{d+1} x \, \Lcal(D_{\mu} \pi^{i} , \Psi^I, \covD_{\mu}),
\end{align}
where all indices are contracted using the appropriate invariant tensors associated with the unbroken group $H$.

\subsection{Gauge Symmetries} \label{sec:gauge coset}

One can also extend the previous construction to the case in which some or all the symmetries are gauged. This requires the introduction of a gauge field $A_\mu = \vec{A}_\mu \cdot \vec Q$, where the generators $\vec Q$  stand for some or all the $T_A$'s and $X_i$'s. The gauged Maurer-Cartan form then reads:
\begin{align}
    \bfomega = \Omega^{-1}\left(\partial_\mu +i A_\mu \right)\Omega \, {\rm d}  x^\mu \ ,
\end{align}
where $A_{\mu}\longrightarrow g(x)\left (A_{\mu}-i\partial_{\mu}\right )g^{-1}(x)$ under a gauge transformation, as usual. The various components of $\bfomega$ defined in Eq. \eqref{MCForm} transform as before, except that now the building blocks $\covd_{\alpha} \pi^{i}$ and $\Acal_\mu^B$ also depend on the gauge field $A_{\mu}$. 

The gauged coset construction is, of course, natural to consider in the context of spontaneously broken gauge theories, in which case $A_\mu$ is a dynamical field. Even in theories with spontaneously broken global symmetries, though, it can be fruitful to introduce non-dynamical gauge fields to aid in the path-integral description of such systems. More precisely, treating the $A_\mu$ as an external field, one can define a generating functional for correlators of the conserved currents $\mathcal{J}^\mu$ associated with the symmetries that have been gauged, as in:
\begin{align}
Z[A]&=\int\Dcal\pivec\, e^{iS_{\rm gauged}}\  , \qquad i^n \frac{\delta^{n}Z[A]}{\delta A_{\mu_{1}}(x_{1})\ldots \delta A_{\mu_{n}}(x_{n})}\Big|_{A=0}\equiv \langle T\mathcal{J}^{\mu_{1}}(x_{1})\ldots \mathcal{J}^{\mu_{n}}(x_{n})\rangle .\label{InOutGaugedGenerator}
\end{align}
It is straightforward to show that $Z[A]$ is gauge-invariant (in the absence of anomalies), from which the conservation of the corresponding currents $\mathcal{J}^{\mu}$ (and Ward-identities, more generally) follows. See e.g. \cite{Leutwyler:1993iq,Leutwyler:1993gf,Glorioso:2020loc} for reviews of this construction.

\section{Schwinger-Keldysh Effective Actions from a Coset Construction} 
\label{sec:coset in-in}

We will now discuss how the coset construction reviewed in the previous section can also be used to construct Schwinger-Keldysh effective actions. In Sec. \ref{sec:symmetries} we argued that, if the single-field effective action is invariant under the global symmetry group $G$, the corresponding Schwinger-Keldysh effective action for a closed system must enjoy twice as many symmetries---i.e. be invariant under the group $G_1 \times G_2$---in a regime where the action is local. How many of these symmetries are realized linearly will depend on the state of the system $\rho$.  Inspired by the form of the  generating functional in Eq. \eqref{ConnectedGeneratorPIInsertions}, we will \emph{postulate} that the symmetry breaking pattern should be determined by acting with a $G_1 \times G_2$ transformation on the state $\rho$ as follows:\footnote{We should stress that Eq. \eqref{in-in rho transformation} is not in contradiction with the familiar statement that symmetry transformations must act on density matrices as $\rho \to U \rho \, U^{-1}$ in order to preserve the trace. When the state of the system is encoded using a local Schwinger-Keldysh effective action, one may wonder what part of $G_1 \times G_2$ is preserved by the ground state in the Hilbert space of this effective theory. Eq. \eqref{in-in rho transformation} provides a way of answering this question. We will support the validity of this criterion in the hydrodynamic regime by discussing a series of non-trivial examples in Sec. \ref{sec:Examples}. Note however that this criterion does not apply to free fields in their ground state---see App. \ref{app:FreeInIn}---for which a hydrodynamic regime does not exist.}
\begin{equation}
	\rho \, \to \, U_1 \rho \, U_2^{-1}, \label{in-in rho transformation}
\end{equation}
where $U_1$ $(U_2)$ is an element of $G_1$ $(G_2)$. Symmetry transformations of this form that do not leave $\rho$ invariant are spontaneously broken. As we will see, this rule of thumb implies different symmetry breaking patterns depending on whether the state $\rho$ is pure or thermal.

\subsection{Closed Systems in a Pure State}

Let us start by considering the case of pure states: $\rho =|\psi \rangle \langle \psi |$. Spontaneous symmetry breaking occurs if there exists a local operator $\Ocal$ whose expectation value on $|\psi \rangle$ is not invariant under some symmetry transformations. More precisely, the symmetry generator $X$ is spontaneously broken if
\begin{align}
	\text{Tr} (\rho [ X, \Ocal]) = \langle \psi | [ X, \Ocal] |\psi \rangle \neq 0.
\end{align}
This condition can be satisfied only if the state $|\psi \rangle$ is an eigenstate of some but not all of the generators of the symmetry group $G$---say, those spanning a subgroup $H$. As a result, the state $\rho$ remains invariant under a transformation \eqref{in-in rho transformation} only when $U_1 \in H_1$  and $U_2 \in H_2$. In other words, the pure state $\rho$ effectively gives rise to a symmetry breaking pattern $G_1 \times G_2 \to H_1 \times H_2$. Denoting with $X_1^i \, (X_2^i)$ the broken generators of $G_1 \, (G_2)$, we introduce for later convenience the linear combinations of generators 
\begin{align}
	X_{r}^i  = X_{1}^i +  X_{2}^i , \qquad\quad   X_{a}^i =  X_{1}^i -  X_{2}^i .
\end{align}
Then, the coset parametrization in the case of pure states can be chosen to be of the form
\begin{tcolorbox}[colframe=white,arc=0pt,colback=greyish2]
\vspace{-8pt}
\begin{align}
	\Omega= e^{i\pi_{r}\cdot X_{r}}e^{i\pi_{a}\cdot X_{a}} \qquad \qquad \quad \text{(closed system in a pure state)}  .
\end{align}
\vspace{-11pt}
\end{tcolorbox}
\noindent For the purposes of calculating the corresponding Maurer-Cartan form, it is important to keep in mind that, while $[X_{1}^i,  X_{2}^j] = 0$, the $X_{r}$'s and $X_{a}$'s do not commute with each other. Let us now turn our attention to thermal states.

\subsection{Closed Systems in a Thermal State}

It would appear at first sight that the matrix elements of a thermal state $\rho \sim e^{- \mathcal{H}/T}$ should always be invariant under the diagonal subgroup of $G_1 \times G_2$ based on the very definition of symmetry, {\it i.e.}  $U \mathcal{H} \, U^{-1} = \mathcal{H}$. This conclusion however would be too hasty~\cite{takahashi1996thermo,Sieberer:2015svu,Beekman:2019pmi}. In order to formulate more precisely the criterion for SSB around a thermal state, we can follow \cite{ForsterBook} and deform the Hamiltonian by adding an operator $\Delta \mathcal{H}$ that explicitly breaks the global symmetry under consideration: $\mathcal{H} \to \mathcal{H} + \Delta \mathcal{H}$. We will denote the resulting canonical ensemble with $\rho_{\Delta \mathcal{H}}$. Then, SSB occurs whenever there is an order parameter $\Ocal$ whose expectation value is not invariant even in the limit $\Delta H \to 0$. Denoting a symmetry transformation as $U=e^{i\alpha X}$ with $X$ a symmetry generator, this is equivalent to the statement that 
    \begin{align}
	    \lim_{\Delta \mathcal{H} \to 0} \tr \! \left( \rho_{\Delta \mathcal{H}}  [ X, \Ocal ] \right) \neq 0 . \label{finite T SSB}
    \end{align}

We can now diagonalize the maximum number of generators of $G$ that commute simultaneously with each other and with the modified Hamiltonian $\mathcal{H} + \Delta \mathcal{H}$, and write the thermal state more explicitly as\footnote{Working only with eigenstates of those generators that commute with  $\mathcal{H} + \Delta \mathcal{H}$ rather than $\mathcal{H}$ alone is formally equivalent to modifying $e^{- \mathcal{H}/T}$ by adding a projection operator~\cite{ForsterBook}.}
\begin{align}
	\rho \sim \sum_{E, \vec Q,\alpha} e^{- E/T} | E, \vec Q, \alpha \rangle \langle E, \vec Q, \alpha|,
\end{align}
where $E$ is the eigenvalue of $\mathcal{H} + \Delta \mathcal{H}$, $\vec Q$ are the charges associated with the commuting generators, and $\alpha$ is an additional collective index that accounts for all possible degeneracies. The states $|E, \vec Q, \alpha \rangle$ form all possible representations of the unbroken group $H$, and when we act on the thermal state with an $H_1 \times H_2$ transformation as in \eqref{in-in rho transformation}, only the elements of the diagonal subgroup $H_{\rm diag}$ will leave $\rho$ invariant. 

It might be helpful to illustrate this point with a simple example. Consider a system where an internal $SO(4)$ is spontaneously broken down to $SO(3)$. Then, the deformation $\Delta \mathcal{H}$ breaks $SO(4)$ explicitly down to $SO(3)$. We can diagonalize at most one generator of $SO(3)$ together with $\mathcal{H} + \Delta \mathcal{H}$, and as a result the thermal state takes the form
\begin{align} \label{thermal state explicit}
	\rho \sim \sum_{E, m, j} e^{- E/T} | E, m, j \rangle \langle E, m, j |,
\end{align}
with $m$ and $j$ the usual $SO(3)$ quantum numbers, playing respectively the role of charge $Q$ and degeneracy parameter $\alpha$. Let us now act with a $SO(3)_1 \times SO(3)_2$ transformation on Eq. \eqref{thermal state explicit}, and we obtain
\begin{align}
	\rho \to & \sum_{E, m, j} \sum_{m',m''} e^{- E/T} D^{(j)}_{m',m} (U_1) | E, m', j \rangle \langle E, m'', j | D^{(j) \, *}_{m'',m} (U_2) \nonumber \\
	& =  \sum_{E, m', m'', j}  e^{- E/T} D^{(j)}_{m',m''} (U_1 U_2^{-1}) | E, m', j \rangle \langle E, m'', j |, \label{rotation SO(3) example}
\end{align}
where we obtained the second line by using standard properties of the $D^{(j)}_{m',m}$ matrices, namely $D^{(j)}_{m',m} (U^{-1}) = D^{(j)\, *}_{m,m'} (U)$ and $\sum_m D^{(j)}_{m',m} (U_1) D^{(j)}_{m,m''} (U_2) = D^{(j)}_{m',m''} (U_1 U_2)$~\cite{Sakurai:2011zz}. Eq. \eqref{rotation SO(3) example} shows that $\rho$ is only invariant under the diagonal subgroup of  $SO(3)_1 \times SO(3)_2$, i.e. when $U_1 = U_2$, while the off-diagonal combination of generators is spontaneously broken.

We conclude therefore that a thermal state realizes the symmetry breaking pattern $G_1 \times G_2 \to H_{\rm diag}$, which differs from the pattern of a pure state. Denoting with $T_1^A \, (T_2^A)$ the generators of $H_1 \, (H_2)$, we introduce the linear combinations of generators 
\begin{align}
	T_{r}^A  = T_{1}^A +  T_{2}^A , \qquad\quad   T_{a}^A =  T_{1}^A -  T_{2}^A ,
\end{align}
with $T_{r}^A$ the generators of $H_{\rm diag}$. We are therefore led to the conclusion that the appropriate coset parametrization for a thermal state should be
\begin{tcolorbox}[colframe=white,arc=0pt,colback=greyish2]
\vspace{-8pt}
\begin{align}
	\Omega= e^{i\pi_{r}\cdot X_{r}}e^{i\pi_{a}\cdot X_{a}} e^{i\varphi_{a}\cdot T_{a}} ,  \label{coset thermal}  \qquad \qquad \quad \text{(closed system in a thermal state),} 
\end{align}
\vspace{-11pt}
\end{tcolorbox}
\noindent where the specific ordering of the various factors has been chosen for later convenience. However, the field content in Eq. \eqref{coset thermal} is by itself incompatible with the principle that Schwinger-Keldysh effective actions should contain twice the number of degrees of freedom as regular in-out effective actions---see discussion in Sec. \ref{sec: field content}. In other words, the fields $\varphi_{a}$ are missing their ``$r$'' partners which, among other things, are necessary to implement the DKMS symmetry.

To remedy this situation, we will add to our field content a set of ``matter fields'' $\rho_r^A$ in the adjoint representation of the unbroken group $H_{\rm diag}$. In accordance with the standard rules of the coset construction~\cite{Weinberg:1996kr}, these matter fields will transform under a generic transformation $g \in G_1 \times G_2$~as
\begin{align}
	\rho_r^A \to h(g, \pi_r, \pi_a, \varphi_a, x)^A{}_B \, \rho_r^B ,
\end{align}
with $h$ some element of $H_{\rm diag}$. As we will see, these fields admit a simple physical interpretation: in the classical limit, they are related to the densities of unbroken charges.

\subsection{Implementing the dynamical KMS symmetry} \label{sec: DKMS coset}

When a system is in a thermal state, correlators must satisfy the KMS condition. This, in turn, imposes some restrictions on the form of the Schwinger-Keldysh effective action. When all the symmetries are linearly realized, these restrictions are enforced by invariance under a DKMS symmetry transformation of the form \eqref{DKMS transformations}. We will now discuss how this symmetry should be implemented on the Goldstone fields and the matter fields $\rho_r$.

To this end, it is helpful to discuss how the KMS condition would affect, say, the 2-point correlation functions of the conserved Noether currents in the effective theory: 
\begin{align}
\langle \mathcal{J}_\mu(t, \vec x) \mathcal{J}_\nu(t', \vec{x}^{\, \prime}) \rangle = \langle  \mathcal{J}_\nu(t'- i \beta/2, \vec{x}^{\, \prime}) \mathcal{J}_\mu(t + i \beta/2, \vec x) \rangle , \label{KMS condition Noether currents}
\end{align}
where we have suppressed the index labeling the corresponding symmetry generators to streamline the notation. We can introduce a generating functional similar to the one in Eq. \eqref{ConnectedGeneratorFromTrace} that allows us to systematically calculate such correlators. In this case, the external sources for the $\mathcal{J}_\mu$'s are gauge fields $A_{1,\mu}$ and $A_{2,\mu}$, and the KMS condition \eqref{KMS condition Noether currents} implies that the generating functional must satisfy the following property:\footnote{We are implicitly assuming here that DKMS transformations are implemented without resorting to additional discrete symmetries---see comment in footnote \ref{footnote: general DKMS}. Eq. \eqref{DKMSGaugedGenerator} could be generalized by allowing such discrete symmetries to act on the gauge fields on the right-hand side. It is easy to work out how the following results would need to be modified.}
\begin{align}
    Z[A_{1}(t), A_{2}(t)] = Z[A_{1}(-t+i\beta/2), A_{2}(-t- i\beta/2)]\ .\label{DKMSGaugedGenerator}
\end{align}

Switching to the $(a,r)$ basis of fields, and expanding in powers of $E/T$, this means that $Z$ should be invariant under the following transformation of sources:  
\begin{subequations} \label{DKMS external gauge sources}
    \begin{align}
        A_{r}(t) &\to A_{r}(-t)-\frac{i\beta}{4}\partial_{t}A_{a}(-t)+\Ocal(E^2/T^2)\ , \\
        A_{a}(t)& \to A_{a}(-t)-i\beta\partial_{t}A_{r}(-t)+\Ocal(E^2/T^2) .
    \end{align}
\end{subequations}
Within the context of the coset construction, external gauge fields can be introduced by gauging the Maurer-Cartan form as discussed in Sec. \ref{sec:gauge coset}:
\begin{align} \label{eq: MC form finite temperature 2}
	&\Omega^{-1}\left(\partial_\mu +i A_{\mu}^{r} \cdot  X_r +i A_{\mu}^{a} \cdot X_a  +i A_{\mu}^{a} \cdot T_a  +i A_{\mu}^{r} \cdot T_r  \right)\Omega \\
	& \qquad \qquad \qquad \qquad \qquad \qquad \qquad \equiv i  \left(  D_\mu \pi_r \cdot X_r +   D_\mu \pi_a \cdot X_a +  D_\mu \varphi_a \cdot T_a + \mathcal{A}_\mu \cdot T_r \right)   \nonumber ,
\end{align}
where we have denoted with a dot the contraction of indices labeling the various generators. The coset covariant derivatives and connection now depend both on the Goldstone fields and the external gauge fields. 

Our goal is to derive how the DKMS symmetry should be implemented on the Goldstone fields and the matter fields $\rho_r$ to ensure  that the generating functional is invariant under \eqref{DKMS external gauge sources}. To this end, we will start by considering a situation where all the generators are spontaneously broken. This is not the physical symmetry breaking pattern we are actually interested in: it is just a convenient trick we will use to figure out how our Goldstone and matter fields should transform. If all the symmetries were spontaneously broken, the coset parametrization would read 
\begin{align}
    \tilde \Omega = e^{i\pi_{r}\cdot X_{r}}e^{i\pi_{a}\cdot X_{a}} e^{i\varphi_{a}\cdot T_{a}}	 e^{i\varphi_{r}\cdot T_{r}} \ ,
\end{align}
and the components of the associated Maurer-Cartan form would be 
\begin{align} \label{eq: tilde MC form finite temperature}
   \tilde \Omega^{-1} &\left(\partial_\mu +i A_{\mu}^{r} \cdot  X_r +i A_{\mu}^{a} \cdot X_a  +i A_{\mu}^{a} \cdot T_a  +i A_{\mu}^{r} \cdot T_r  \right) \tilde \Omega \nonumber \\ &\equiv i \, ( \tilde D_\mu \pi_r \cdot X_r + \tilde D_\mu \pi_a \cdot X_a +  \tilde D_\mu \varphi_a \cdot T_a + \tilde D_\mu \varphi_r \cdot T_r ) \\
   & = i \,  e^{- i\varphi_{r}\cdot T_{r}} \left[ D_\mu \pi_r \cdot X_r +  D_\mu \pi_a \cdot X_a +  D_\mu \varphi_a \cdot T_a + \left( i  e^{i\varphi_{r}\cdot T_{r}}  \partial_\mu  e^{- i\varphi_{r}\cdot T_{r}}  + \mathcal{A}_\mu \right) \cdot T_r \right]  e^{i\varphi_{r}\cdot T_{r}} \ . \nonumber
\end{align}
In the second line, we are showing explicitly how the building blocks of our new coset are related to the ``physical'' ones defined in Eq. \eqref{eq: MC form finite temperature 2}. 

The advantage of considering an auxiliary symmetry breaking pattern where all the symmetries are spontaneously broken is twofold: first, all symmetries are now treated on equal footing and, in particular, all components of our external gauge fields can be obtained from a covariant derivative by turning off the Goldstone fields: 
\begin{align}
	    \tilde D_\mu \pi^i_r \to A^{ri}_{\mu} , \qquad \quad 
	    \tilde D_\mu \pi^i_a \to A^{ai}_{\mu}, \qquad \quad 
	    \tilde D_\mu \varphi^B_a \to A^{aB}_{\mu}, \qquad \quad 
	    \tilde D_\mu \varphi^B_r \to A^{rB}_{\mu} \ .
\end{align}
Therefore, if this was the symmetry breaking pattern that we were interested in, we could ensure that the generating functional is invariant under the transformations \eqref{DKMS external gauge sources} by demanding that the effective action be symmetric under\footnote{An important comment on our notation: when $\mu = t$, equations \eqref{eq: DKMS tilde derivatives} reduce to $\tilde D_t \pi_r (t) \to \tilde D_{t} \pi_r (-t) + ... = - \tilde D_{-t} \pi_r (-t) + ... $, and so on. The same goes for Eqs. \eqref{eq: DKMS derivatives all broken}.}
\begin{subequations} \label{eq: DKMS tilde derivatives}
    \begin{align}
        \tilde D_\mu \pi_a (t) \to  \tilde D_\mu \pi_a  (-t) - i \beta \partial_t  \tilde D_\mu \pi_r (-t) +\Ocal(E^2/T^2)\ , \\
        \tilde D_\mu \pi_r (t) \to  \tilde D_\mu \pi_r  (-t) - \frac{i \beta}{4} \partial_t  \tilde D_\mu \pi_a (-t) +\Ocal(E^2/T^2)\ ,  \\
        \tilde D_\mu \varphi_a (t) \to  \tilde D_\mu \varphi_a  (-t) - i \beta \partial_t  \tilde D_\mu \varphi_r (-t) +\Ocal(E^2/T^2)\ , \\
        \tilde D_\mu \varphi_r (t) \to  \tilde D_\mu \varphi_r  (-t) - \frac{i \beta}{4} \partial_t  \tilde D_\mu \varphi_a (-t) +\Ocal(E^2/T^2)\ .
    \end{align}
\end{subequations}
The second advantage is that the coset connection is now trivial, and therefore higher covariant derivatives of the Goldstone fields can be obtained by acting with regular partial derivatives on the Maurer-Cartan components; hence, the symmetry transformations \eqref{eq: DKMS tilde derivatives} are covariant under all the non-linearly realized symmetries. 

Based on the second line of \eqref{eq: tilde MC form finite temperature}, they can be expressed equivalently in terms of the physical covariant derivatives $\mathcal{D}_\mu \pi_r, \mathcal{D}_\mu \pi_a, \mathcal{D}_\mu \varphi_a, \nabla_\mu$ and the combination 
\begin{align}
	D_\mu \varphi_r \equiv i  \, e^{i\varphi_{r}\cdot T_{r}}  \partial_\mu  e^{- i\varphi_{r}\cdot T_{r}}  + \mathcal{A}_\mu 
\end{align}
as follows:\footnote{To derive these expressions, we assumed that the structure constants are totally antisymmetric, which is ensured whenever the symmetry group is compact~\cite{Weinberg:1996kr}.}
\begin{subequations} \label{eq: DKMS derivatives all broken}
    \begin{align}
       & D_\mu \pi_a (t) \to  D_\mu \pi_a  (-t) - i \beta \nabla_t  D_\mu \pi_r (-t) -  \beta \left[ D_t \varphi_r (-t) , D_\mu \pi_r (-t)  \right] + \Ocal(E^2/T^2)\ , \\
       & D_\mu \pi_r (t) \to  D_\mu \pi_r  (-t) - \frac{i \beta}{4} \nabla_t D_\mu \pi_a (-t) - \frac{\beta}{4} \left[ D_t \varphi_r (-t) , D_\mu \pi_a (-t)  \right]+\Ocal(E^2/T^2)\ , \\
       & D_\mu \varphi_a (t) \to  D_\mu \varphi_a  (-t) - i \beta \nabla_t  D_\mu \varphi_r (-t) - \beta \left[ D_t \varphi_r (-t) , D_\mu \varphi_r (-t) \right] + \Ocal(E^2/T^2) \ , \label{eq: last DKMS derivatives all broken} \\
       & D_\mu \varphi_r (t) \to  D_\mu \varphi_r  (-t) - \frac{i \beta}{4} \nabla_t  D_\mu \varphi_a (-t) - \frac{ \beta}{4} \left[ D_t \varphi_r (-t) , D_\mu \varphi_a (-t)  \right] + \Ocal(E^2/T^2)\ .  \label{eq: laster DKMS derivatives all broken}
    \end{align}
\end{subequations}
where, on the right hand side of these equations, $\nabla_t = \partial_t + i \mathcal{A}_t^B (-t) T_B$, and we have streamlined our notation by defining commutators between covariant derivatives, e.g.
\begin{align}
	 \left[ D_t \varphi_r (-t) , D_\mu \pi_a (-t)  \right]_k \equiv i f_{Ajk} D_t \varphi_r^A (-t)  D_\mu \pi_a^j (-t) \ ,
\end{align}
and similarly for the other commutators.\footnote{Note that all four components $X_r,X_a,T_a$ and $T_r$ are in independent representations of $T_r$ and therefore do not mix. This is due to the structure of commutation relations, e.g.the term $e^{-i\varphi_r \cdot T_r} D_\mu \pi_r \cdot X_r e^{i \varphi_r \cdot T_r}$ is a linear combination of $X_r$ generators due to $[X_r,T_r] \sim X_r$. For this reason, \eqref{eq: DKMS tilde derivatives} is related to \eqref{eq: DKMS derivatives all broken} through conjugation with $e^{-i \varphi_r \cdot T_r}$} Once again, the transformations \eqref{eq: DKMS derivatives all broken} are manifestly covariant under all the non-linearly realized symmetries. 

At this point, we notice that the quantity $D_t \varphi_r$ has exactly the same transformation properties as our matter fields $\rho_r$ (we are focusing on internal symmetries, and therefore boosts are irrelevant---i.e. explicitly broken---as far as we are concerned). Therefore, if Eqs. \eqref{eq: DKMS derivatives all broken} involved only $D_t \varphi_r$ and its derivatives, we could simply replace $D_t \varphi_r (t)\to \rho_r(t)$ everywhere and obtain  DKMS symmetry transformations involving the Goldstones $\pi_r, \pi_a, \varphi_a$ and the matter fields $\rho_r$. Unfortunately, Eq. \eqref{eq: last DKMS derivatives all broken} depends on all components of $D_\mu \varphi_r$, but this can be remedied by ``commuting'' the covariant derivatives of $\nabla_t  D_\mu \varphi_r (-t)$: tedious but straightforward manipulations show that 
\begin{align}
	\nabla_t  D_\mu \varphi_r (-t) - i [ D_t \varphi_r (-t), D_\mu \varphi_r (-t) ] = \mathcal{F}_{t \mu} (-t) + \nabla_\mu D_t \varphi_r (-t) \ ,
\end{align}
where $\mathcal{F}_{\mu\nu}$ is the usual field strength tensor associated with $\mathcal{A}_\mu$. Using this identity, and replacing $D_t \varphi_r (t) \to \rho_r(t) $ everywhere,\footnote{This also means replacing $D_t \varphi_r (-t) = - D_{-t} \varphi_r (-t) \to - \rho_r (-t)$.}
we finally obtain the desired form of the DKMS transformation rules for our Goldstone and matter fields (we only need to consider the $\mu = t$ component of Eq. \eqref{eq: laster DKMS derivatives all broken}):
\begin{tcolorbox}[colframe=white,arc=0pt,colback=greyish2]
\vspace{-8pt}
\begin{subequations} \label{eq: DKMS transformations with matter fields}
    \begin{align}
       & D_\mu \pi_a (t) \to  D_\mu \pi_a  (-t) - i \beta \nabla_t  D_\mu \pi_r (-t) +  \beta \left[ \rho_r (-t) , D_\mu \pi_r (-t)  \right] + \Ocal(E^2/T^2)\ , \\
       & D_\mu \pi_r (t) \to  D_\mu \pi_r  (-t) - \frac{i \beta}{4} \nabla_t D_\mu \pi_a (-t) + \frac{\beta}{4} \left[ \rho_r (-t) , D_\mu \pi_a (-t)  \right]+\Ocal(E^2/T^2)\ , \\
       & D_\mu \varphi_a (t) \to  D_\mu \varphi_a  (-t) + i \beta \nabla_\mu \rho_r (-t) -i \beta \mathcal{F}_{t \mu} (-t) +  \Ocal(E^2/T^2) \ , \\
       & \rho_r (t) \to  - \rho_r  (-t) - \frac{i \beta}{4} \nabla_t  D_t \varphi_a (-t) + \frac{ \beta}{4} \left[ \rho_r (-t) , D_t \varphi_a (-t)  \right] + \Ocal(E^2/T^2)\ .
    \end{align}
\end{subequations}
\vspace{-11pt}
\end{tcolorbox}
\noindent These transformation rules are the main result of this subsection. They act non-locally at the level of the fields, but to the best of our knowledge there is no fundamental obstruction to having non-local discrete symmetries. Furthermore, because the Goldstone fields enter the effective action only through their covariant derivatives, and these transform locally, it is not hard to impose the DKMS symmetry in practice. In this paper, we will be concerned with the implementation of DKMS relations up to first order in $E/T$; the systematics of higher order corrections are still an open question that we hope to explore in the near future. We will derive the lowest-order invariant combinations in Sec. \eqref{sec:invariatn combinations}.

\subsection{Relation to other approaches in the literature}

We should briefly comment on the relation between our approach and previous results in the literature on out-of-equilibrium effective actions. It was previously proposed that, in the hydrodynamic limit, the Schwinger-Keldysh effective action should contain one Goldstone field for each continuous symmetry, regardless of whether it is spontaneously broken or not (see e.g.~\cite{Glorioso:2018wxw, Landry:2019iel,Glorioso:2020loc}). The effective action then must be invariant under an additional local symmetry, which only depends on the spatial coordinates and acts on the fields associated with unbroken generators as follows: 
\begin{align} \label{eq: chemical shift}
    e^{i\varphi_r (t, \vec x) \cdot T_r} \quad \to \quad e^{i\varphi_r (t, \vec x) \cdot T_r} h_r(\vec x) \ , \, \qquad \qquad \qquad  h_r(\vec x) \equiv e^{i c (\vec x) \cdot T_r} \ .	
\end{align}
In the simplest, abelian case, this symmetry reduces to a local shift, $\varphi_r (t, \vec x)  \to \varphi_r  (t, \vec x) + c (\vec x)$. The symmetry \eqref{eq: chemical shift} is equivalent to the transformation $\tilde \Omega \to \tilde \Omega \, h_r(\vec x)$, which in turn implies the following transformation rules for the coefficients of the Maurer-Cartan form: 
\begin{subequations} \label{eq: diffusive symmetry tilde derivatives}
    \begin{align}
        \tilde D_\mu \pi_a \cdot X_a &\to \tilde D_\mu \pi_a \cdot h_r^{-1} (\vec x) X_a h_r (\vec x) \ , \\
	    \tilde D_\mu \pi_r \cdot X_r &\to \tilde D_\mu \pi_r \cdot h_r^{-1} (\vec x) X_r h_r (\vec x) \ , \\
        \tilde D_\mu \varphi_a \cdot T_a &\to \tilde D_\mu \varphi_a \cdot h_r^{-1} (\vec x) T_a h_r (\vec x) \ , \\
        \tilde D_\mu \varphi_r \cdot T_r &\to \tilde D_\mu \varphi_r \cdot h_r^{-1} (\vec x) T_r h_r (\vec x) - i \delta_\mu^j h_r^{-1} (\vec x)  \partial_j h_r (\vec x) \ .
    \end{align}
\end{subequations}
This symmetry plays two important roles: (1) it effectively forces us to contract all the indices in a way that is invariant under $H_{\rm diag}$, even though $H_{\rm diag}$ is formally broken; and (2) it ensures that the fields $\varphi_r$ appear in the effective action only through $\tilde D_t \varphi_r$ and its covariant derivatives---the spatial components $\tilde D_i \varphi_r$'s are not allowed building blocks. This, in turn, gives rise to a diffusive behavior for the unbroken currents, and for this reason we'll also refer to invariance under \eqref{eq: diffusive symmetry tilde derivatives} as \emph{diffusive symmetry}. This additional symmetry is to be regarded as emergent at low energies, and its physical origin is not particularly clear.\footnote{See however~\cite{deBoer:2018qqm,Glorioso:2018mmw} for a holographic interpretation of this symmetry.} Furthermore, its implications have so far been explored mostly in the classical limit, i.e. by working only up to quadratic order in the $a-$type fields. This approach has the advantage that the DKMS symmetry becomes easier to implement~\cite{Glorioso:2017fpd}---which is why we started this section by considering a similar symmetry breaking pattern. However, in practice one is actually more interested in the properties of the charge density $\rho_r$ rather than the field $\varphi_r$, and some authors even resort to an explicit change of variable from the latter to the former---see e.g.~\cite{Delacretaz:2023ypv}, which performs precise numerical tests of the EFT approach to diffusion.

This alternative approach yields exactly the same correlation functions for conserved currents as the one developed in this paper. However, given the different number of time derivatives at play, we don't expect these two approaches to be fully equivalent, and ultimately expect the number of propagating degrees of freedom to be different---at least based on our experience with more conventional EFTs. We plan to further investigate this question in the near future, but in the meantime we find the conceptual simplicity of our approach---which relies on the standard rules of the coset construction and doesn't require additional symmetries---particularly compelling.

\subsection{Lowest order DKMS-invariant combinations} \label{sec:invariatn combinations}

Now that we understand how the DKMS symmetry acts on our Goldstone and matter fields up to first order in $E/T$, we can try to build combinations that are invariant under these symmetries at this order. To this end, it is once again convenient to work at first with our ``fictitious'' covariant derivatives $\tilde D_\mu \pi_r, \dots$ because the DKMS transformation rules \eqref{eq: DKMS tilde derivatives} are simpler than the ones for the physical fields given in \eqref{eq: DKMS transformations with matter fields}. We will eventually rewrite the combinations we derive purely in terms of the physical fields, but to do so we'll need to impose the additional diffusive symmetry introduced in the previous section. 

It is easy to see that the following combination changes by a total derivative up to first order in~$E/T$,
\begin{align}
	\tilde D_t \pi_a \cdot \tilde D_t \pi_r \quad \to \quad \tilde D_t \pi_a \cdot \tilde D_t \pi_r - \frac{i \beta}{2} \partial_t \left[ \tilde D_t \pi_r  \cdot \tilde D_t \pi_r  - \frac{1}{4} \tilde D_t \pi_a \cdot  \tilde D_t \pi_a \right]  +\Ocal(E^2/T^2) ,
\end{align}
and thus provides an invariant contribution to the effective action. The dot in this equation stands for the most general symmetric contraction that is invariant under the diffusive symmetry. An analogous statement can be made about $\tilde D_i \pi_a \cdot \tilde D^i \pi_r$ and $\tilde D_t \varphi_a \cdot \tilde D_t \varphi_r$. Using Eq. \eqref{eq: tilde MC form finite temperature}, it is easy to see that these combinations are respectively equal to $D_t \pi_a \cdot D_t \pi_r$, $ D_i \pi_a \cdot D^i \pi_r$, and $D_t \varphi_a \cdot D_t \varphi_r \to D_t \varphi_a \cdot \rho_r$, which are therefore invariant under the DKMS symmetry up to a total derivative. 

The last combination that would be natural to consider, $\tilde D_i \varphi_a \cdot \tilde D^i \varphi_r$, would also be invariant under DKMS but, alas, not under the diffusive symmetry. This suggests that we should act with at least one time derivative on $\tilde D^i \varphi_r$, but the contraction $\tilde D_i \varphi_a \cdot \partial_t \tilde D^i \varphi_r$, albeit now invariant under the diffusive symmetry, would now no longer be invariant under DKMS. However, this can be easily remedied by adding a term quadratic in $\tilde D_i \varphi_a$ to form the combination
\begin{align}
	\tilde D^j \varphi_a \cdot \left(\partial_t \tilde D_j \varphi_r - \frac{i}{\beta} \tilde D_j \varphi_a \right) \ , 
\end{align}
which is exactly invariant under a DKMS transformation up to $\Ocal(E^2/T^2)$. Performing manipulations analogous to those that took us from Eqs. \eqref{eq: DKMS tilde derivatives} to Eqs. \eqref{eq: DKMS transformations with matter fields}, we can rewrite this combination in terms of our ``physical'' fields as follows:
\begin{align}
	D^j \varphi_a \cdot \left(\nabla_j \rho_r + \mathcal{F}_{tj} - \frac{i}{\beta} D_j \varphi_a \right) \ .
\end{align}
The combinations we have identified above provide the leading kinetic terms for our fields. Of course, the DKMS symmetry can also be imposed at higher orders. Examples of such higher order terms that will play a role in our discussion of antiferromagnets (see Sec. \ref{sec:Antiferromagnets}) are 
\begin{align}
	D^j \pi_{a}\cdot \left (-\covD_{t} D_j\pi_{r}+i\left [\rho_{r},D_{j}\pi_{r}\right ]+\frac{i}{\beta}D_j\pi_{a}\right ) \ , 
\end{align}
or an equivalent expression with spatial derivatives replaced by time derivatives.

\subsection{Power counting rules}

As we discussed in Sec. \ref{sec:locality}, in order for our EFT to be well defined we must be able to assign to each term in the effective action a definite scaling in terms of our expansion parameters. This, in turn, requires us to specify how covariant derivatives and matter fields scale with energy and momentum. The scaling of covariant derivatives is the conventional one---time derivatives scale like energy, spatial derivatives like momentum. The scaling of matter fields, instead, is determined by the requirement that the DKMS transformations \eqref{eq: DKMS transformations with matter fields} are organized in an expansion of $E/T$. As a result, we have that
\begin{tcolorbox}[colframe=white,arc=0pt,colback=greyish2]
\vspace{-8pt}
\begin{align} \label{eq: scaling rules}
   \rho_r, \, D_t \pi_r, \, D_t \pi_a, \, D_t \varphi_a, \, \nabla_t \, \sim E, \qquad \qquad \quad D_i \pi_r, \, D_i \pi_a, \, D_i \varphi_a, \, \nabla_i \, \sim k.
\end{align}
\vspace{-11pt}
\end{tcolorbox}
\noindent In the next section we will discuss a few concrete examples, and show how these power counting rules can be used to estimate the size of various operators in the effective action.

\newpage

\section{Examples: Paramagnets, Antiferromagnets, and Ferromagnets} \label{sec:Examples}

 Non-relativistic magnetic systems at finite temperature are endowed with an internal $SO(3)$ symmetry that corresponds to global rotations of all the spins, and thus provide a non-trivial testing ground for our formalism. In the case of paramagnets this $SO(3)$ symmetry remains unbroken, while in (anti-)ferromagnets it is spontaneously broken down to an $SO(2)$ subgroup. The feature that sets ferromagnets apart is that they have a non-zero density of unbroken $SO(2)$ charge---i.e., a non-zero magnetization density. In this section, we will discuss separately these three possibilities, restricting our attention to the internal $SO(3)$ symmetry and neglect the space-time symmetries that would also be broken by a finite temperature state.\footnote{This amounts to neglecting phonon excitations by working in the incompressible limit and, in particular, treating boosts as if they were explicitly broken. An EFT treatment of the Goldstone modes arising at zero temperature from the simultaneous breaking of $SO(3)$ and spacetime symmetries was recently discussed in~\cite{Pavaskar:2021pfo}.} These examples will illustrate how to power count terms in the effective action and how to calculate correlation functions of Noether currents.\footnote{Classic studies of the high-$T$ Noether current correlator for these systems include for instance \cite{Halperin:1969wrj,ForsterBook}.}

\subsection{Paramagnets\label{sec:Paramagnets}}

Paramagnets are systems where the internal $SO(3)$ symmetry remains unbroken.\footnote{See \cite{Glorioso:2020loc} for a more general study of non-Abelian hydrodynamics in the absence of SSB and in the classical limit.} As a result, the fields $\pi_r$ and $\pi_a$ are absent, the only fields that enter the low energy EFT are the triplet of Goldstone fields $\vec \varphi_a$ corresponding to the breaking of $SO(3)_1 \times SO(3)_2$ down to the diagonal subgroup $SO(3)_{\rm diag}$, and the three associated matter fields $\vec \rho_r$. Using the invariant building blocks we have identified in Sec. \ref{sec:invariatn combinations}, we can write the following leading-order effective action for a paramagnet:
\begin{tcolorbox}[colframe=white,arc=0pt,colback=greyish2]
\vspace{-8pt}
\begin{align} \label{eq: S para}
	S_{\rm para}  = \int dt \, d^3 x \left[ \frac{k_*^3}{E_* } \, \rho_r \cdot D_t \varphi_a  - k_* \, D^j\varphi_{a}  \cdot \left(\nabla_{j}\rho_{r} + \mathcal{F}_{tj}^{r} - \frac{i}{\beta}D_{j} \varphi_{a}\right) \right] \ ,
\end{align}
\vspace{-11pt}
\end{tcolorbox}
\noindent where the dot  stands for a contraction of the internal indices with a 3-dimensional Kronecker delta. Note that this action is not invariant under time reversal since this is never a symmetry of Schwinger-Keldysh effective actions, as we have discussed alread in Sec.~\ref{sec:symmetries}. Physically, this makes it possible to reproduce a diffusive behavior, as we will see below.

We have chosen to parametrize the two free coefficients in the effective action \eqref{eq: S para} in terms of some microscopic momentum and energy scales, denoted respectively with $k_*$ and $E_*$, so that our effective action will be organized in powers of $k/k_*$, $E/E_*$, and $E/T$. In fact, there are only two independent expansion parameters, because the first two ratios are related to each other by the free equations of motion for $\rho_r$, which can be obtained by varying the action with respect to $\varphi_a$:\footnote{In the simplest case where $E_* = T$, then there is just one independent expansion parameter.}
\begin{align} \label{eq: eom para}
	\left. \frac{\delta S_{\rm para}}{\delta \varphi_a} \right|_{\varphi_a = A_a = A_r = 0} = - \frac{k_*^3}{E_* }  \partial_t \rho_r + k_* \partial_j\partial^j \rho_r = 0 \ , \qquad \rightarrow \qquad \frac{E}{E_*} \sim \frac{k^2}{k_*^2} \ .
\end{align}
Combining this relation with the scaling rules \eqref{eq: scaling rules} and the fact that $d^4 x \sim E^{-1} k^{-3}$, we  see that the effective action \eqref{eq: S para} contains all the terms of $\mathcal{O}(k_*/k)$, with the term quadratic in $\varphi_a$ further enhanced by a factor of $T/E$. In other words, our action is accurate up to leading order in $k/k_*$, and up to the first subleading order in $E/T$. Notice also that our expression in \eqref{eq: S para} does not rely on a classical approximation: the covariant derivative $D_t \varphi_a$ is generically a non-linear combination of the fields $\varphi_a$, and as such contains terms of all orders in $\hbar$ (see discussion in Sec.~\ref{sec:KeldyshRotation}).

Varying instead our action with respect to the gauge fields $A_\mu^a$ and $A_\mu^r$ yields respectively the conserved currents $\mathcal{J}^\mu_r$ and $\mathcal{J}^\mu_a$ expressed in terms of the external gauge fields, $\rho_r, \varphi_a$, and their derivatives. In particular, setting $A_\mu^a = \varphi_a = 0$, the current $\mathcal{J}_r^\mu$ reduces to the \emph{classical} conserved current in the presence of external gauge fields $A_\mu^r$:
\begin{align} \label{eq: magnetization current}
	\left. \mathcal{J}^\mu_r \right|_{\varphi_a = A_a = 0} = \left(\frac{k_*^3}{E_* } \, \rho_r \ , - k_* \nabla^j \rho_r +k_* F^{tj}_r \right) \ .
\end{align}
From this, we see that the fields $\rho_r$ are equal to the conserved charge densities up to an overall normalization. In the absence of external gauge fields, we recover the standard constitutive relation $ \mathcal{J}_r^i = - \mathcal{D} \partial^i \mathcal{J}_r^0$ with a diffusion coefficient $\mathcal{D} \equiv E_* / k_*^2$. In fact, the equations of motion \eqref{eq: eom para} are just a set of diffusion equations for the charge densities $\mathcal{J}_r^0$. When $\rho_r = 0$, instead, the second term in the classical current densities reproduces Ohm's law, $\mathcal{J}_r^i = \sigma E^i_r$, with $E^i_r$ the electric component of the field strength and $\sigma \equiv k_*$ the conductivity.

More in general, we can calculate correlation functions of the currents $\mathcal{J}^\mu_r$ and $\mathcal{J}^\mu_a$ by taking functional derivatives with respect to $A_\mu^a$ and $A_\mu^r$ of the generating functional 
\begin{align}
	Z [A_\mu^a, A_\mu^r] = \int \mathcal{D} \rho_r \mathcal{D} \varphi_a e^{i S_{\rm para}} \ .
\end{align}
A single derivative with respect to $A_\mu^a$ yields the expectation value of \eqref{eq: magnetization current}, which coincides with the expectation value of the physical Noether current, $\langle \mathcal{J}^\mu \rangle$. The requirement that paramagnets preserve the $SO(3)$ symmetry in the absence of external fields implies that $\langle \rho_r \rangle = 0$.

Similarly, by taking two functional derivatives we can calculate two point functions: 
\begin{align}
	\langle \mathcal{J}^{\mu}_{r A}(p)\mathcal{J}^{\nu}_{r B}(k)\rangle=-\frac{\delta^{2}Z}{\delta A^{A}_{\mu a}(-p) \delta A^{B}_{\nu a}(-k)} \ , \qquad \langle \mathcal{J}^{\mu}_{r A}(p)\mathcal{J}^{\nu}_{a B}(k)\rangle=-\frac{\delta^{2}Z}{\delta A^{A}_{\mu a}(-p) \delta A^{B}_{\nu r}(-k)} \ . \label{JJCorrelatorGeneric}
\end{align}
These current correlators can be inferred from the correlators of $\rho_r$ and $\varphi_a$, and correspond to different Green's functions of the physical Noether current $\mathcal{J}^\mu$, along the lines of what we discussed in Sec. \ref{sec:KeldyshRotation}). For example,
\begin{align}
	\langle \mathcal{J}^\mu_r (x) \mathcal{J}^\nu_r (x') \rangle = \tfrac{1}{2} \langle \{ \mathcal{J}^\mu (x), \mathcal{J}^\nu (x') \} \rangle \ , \qquad  \langle \mathcal{J}^\mu_r (x) \mathcal{J}^\nu_a (x') \rangle = \theta(t-t') \langle [ \mathcal{J}^\mu (x), \mathcal{J}^\nu (x') ] \rangle \ .
\end{align}
For simplicity we will calculate these correlators with vanishing external sources, i.e. setting $A_\mu^r = A_\mu^a = 0$ after taking the appropriate functional derivatives. 

At leading order in $k/k_*$ the current correlators can be calculated by approximating the currents $\mathcal{J}^\mu_r$ and $\mathcal{J}^\mu_a$ up to linear order in the fields  $\rho_r$ and $\varphi_a$; thus, we only need to know the 2-point functions of these fields, which can be obtained simply by inverting the quadratic term in \eqref{eq: S para}---this is in fact one of the main advantages of working with the effective action:
\begin{subequations}
    \begin{align}
        \langle \rho_{r}^{A}(\omega, \vec k)\rho_{r}^{B}(-\omega, -\vec k)\rangle'&= \frac{2}{\beta} \frac{\mathcal{D}}{\sigma} \, \frac{\mathcal{D} k^2\delta^{AB}}{\omega^{2}+\mathcal{D}^2 k^4} \, \\
        \langle \rho_{r}^{A}(\omega, \vec k)\varphi_{a}^{B}(- \omega, -\vec k)\rangle'&= \frac{\mathcal{D}}{\sigma} \,  \frac{\delta^{AB}}{\omega + i \mathcal{D} k^{2}} \ ,  \\
        \langle \varphi_{a}^{A}(\omega, \vec k)\rho_{r}^{B}(-\omega, -\vec k)\rangle'&= \frac{\mathcal{D}}{\sigma} \, \frac{\delta^{AB}}{-\omega + i \mathcal{D} k^{2}}\ ,
    \end{align}
\end{subequations}
where the primes on the left-hand side denote the fact that we have dropped the delta functions imposing energy and momentum conservation. Combining these 2-point functions we can easily calculate all the components of the $r$-$r$ correlator,
\begin{subequations} \label{JJParamagnetrr}
    \begin{align}
    \langle \mathcal{J}^{tA}_{r}(\omega, \vec k)\mathcal{J}^{tB}_{r}(- \omega, -\vec k)\rangle'&=\frac{2\sigma}{\beta}\frac{k^{2}\delta^{AB}}{\omega^{2}+\mathcal{D}^2 k^{4}} \ , \\
    \langle \mathcal{J}^{tA}_{r}(\omega, \vec k)\mathcal{J}^{jB}_{r}(- \omega, -\vec k)\rangle'&=\frac{2\sigma}{\beta}\frac{\omega k^{j}\delta^{AB}}{\omega^{2}+\mathcal{D}^2 k^{4} }\ , \\
    \langle \mathcal{J}^{iA}_{r}(\omega, \vec k)\mathcal{J}^{jB}_{r}(- \omega, -\vec k)\rangle'&=\frac{2\sigma}{\beta}\left (\delta^{ij}- \frac{\mathcal{D}^2 k^{2}k^{i}k^{j}}{\omega^{2}+\mathcal{D}^{2} k^{4}}\right )\delta^{AB}\ ,
    \end{align}
\end{subequations}
which are precisely of the form needed to ensure current conservation, $k^{\mu}\langle J^{A}_{r \mu}J^{B}_{r \nu}\rangle'=0$. 

The retarded $r$-$a$ correlator is computed in a similar way, and we have verified that it is also conserved. For completeness, we list here its components:
\begin{subequations}
\begin{align}
  \langle \mathcal{J}^{tA}_{r}(\omega, \vec k)\mathcal{J}^{tB}_{a}(- \omega, -\vec k)\rangle'&= \frac{\sigma k^{2}\delta^{AB}}{\omega+i\mathcal{D} k^{2}}\label{JJParamagnetratt} \ , \\
  \langle \mathcal{J}^{tA}_{r}(\omega, \vec k)\mathcal{J}^{jB}_{a}(- \omega, -\vec k)\rangle'&= \langle \mathcal{J}^{jA}_{r}(\omega, \vec k)\mathcal{J}^{tB}_{a}(- \omega, -\vec k)\rangle' = \frac{\sigma \omega k^j \delta^{AB}}{\omega + i \mathcal{D} k^2} \ , \\
    \langle \mathcal{J}^{iA}_{r}(\omega, \vec k)\mathcal{J}^{jB}_{a}(- \omega, -\vec k)\rangle'&=\sigma \omega \left( \delta^{ij} - \frac{i \mathcal{D} k^i k^j}{\omega + i \mathcal{D} k^2} \right)\delta^{AB}\ ,
  \end{align}  
\end{subequations}
displaying the usual diffusive pole. All of the above correlators agree with previous well-known results in the literature, see e.g.~\cite{Halperin:1969wrj, ForsterBook}. Note that the correlators above receive contributions from contact terms---i.e. terms in the effective action \eqref{eq: S para} that are quadratic in the external gauge fields---and these are crucial to ensure current conservation.

\subsection{Antiferromagnets\label{sec:Antiferromagnets}}

Antiferromagnets are systems where the internal $SO(3)$ symmetry is spontaneously broken down to $SO(2)$ by a non-trivial staggered-magnetization order parameter which, without loss of generality, we take to be along the 3-direction. Consequently, the relevant degrees of freedom at low energies are the two doublets of Goldstones $\pi_r$ and $\pi_a$, a single Goldstone $\varphi_a$ and its associated matter field $\rho_r$.

The antiferromagnetic ground state differs from the ferromagnetic one in that the magnetization density vanishes. This distinction is often captured by the statement that ferromagnets break time reversal whereas antiferromagnets do not---see e.g.~\cite{Burgess:1998ku,Penco:2020kvy}. This is actually only part of the story, because otherwise the Schwinger-Keldysh effective actions for these two systems would be identical given that time reversal is always broken. The staggered magnetization of the antiferromagnetic ground state picks a preferred direction in spin space but not an orientation. Therefore, it breaks rotations by generic angles around the 1- and 2-directions, but is still invariant under a residual discrete subgroup that consist of 180\textdegree-rotations around these same axes. These discrete symmetries act on our fields as
\begin{align} \label{eq: residual discrete transformations}
	\rho_r \to - \rho_r, \qquad \quad \varphi_a \to - \varphi_a, \qquad \quad \pi_{a, r}^1 \to - \pi_{a, r}^1, \qquad \quad \pi_{a, r}^2 \to \pi_{a, r}^2 ,
\end{align}
in the case of rotations around the 2-axis; for rotations around the 1-axis, it would be the fields $\pi_{a, r}^1$ that are left invariant.\footnote{Rotations around the 2-direction flip the sign of the generators of rotations around the 1- and 3- directions. In our notation, this means that $X^1_{a,r} \to - X^1_{a,r} $ and $T_{a,r} \to - T_{a,r}$. The transformation properties of the Goldstone modes are determined as usual by demanding that coset parametrization $\Omega$ remains invariant when the generators are transformed this way. The transformation rule for $\rho_r$ should be the same as that of its partner field, $\varphi_a$. This is also consistent with that fact that, as we will see, $\rho_r$ ir related to the density of spin along the 3-direction.} The Schwinger-Keldysh effective action for antiferromagnets is invariant under these discrete transformations, whereas the one for ferromagnets is not. This statement applies equally to zero-temperature effective actions. In that context, antiferromagnets are also separately invariant under time reversal, whereas ferromagnets are only invariant under a combination of discrete rotations and time reversal. This state of affairs is summarized in Table \ref{tab: AF vs F}. 

\begin{table}[t]
    \centering
    \begin{tabular}{c|c|c|c|c}
         & {\bf AF $T=0$} & {\bf AF $T\neq 0$} & {\bf F $T=0$} & {\bf F $T\neq 0$} \\
         \hline
         $T $ & \cmark &  \xmark & \xmark  & \xmark \\
          \hline
          $D$ & \cmark  &  \cmark  & \xmark  &\xmark  \\
           \hline
           $T+D$  & \cmark  & \xmark & \cmark  & \xmark \\
            \hline
    \end{tabular}
    \caption{Discrete symmetries ferromagnets (F) and antiferromagnets (AF) in ordinary ($T=0$) and Schwinger-Keldysh ($T\neq 0$) effective actions. We have denoted with $D$ the residual discrete transformations of the form \eqref{eq: residual discrete transformations}. The symbol \cmark (\xmark) indicates unbroken (broken) symmetries.}
    \label{tab: AF vs F}
\end{table}
Our Goldstone covariant derivatives transform in a simple way under the discrete symmetry~\eqref{eq: residual discrete transformations}:
\begin{align}
    D_\mu \varphi_a \to - D_\mu \varphi_a, \qquad \quad  D_\mu \pi_{a,r}^1 \to - D_\mu \pi_{a,r}^1,\qquad \quad  D_\mu \pi_{a,r}^2 \to D_\mu \pi_{a,r}^2 \ .
\end{align}
Keeping also in mind the requirement of DKMS invariance, the low energy effective action for antiferromagnets up to first subleading order in $E/T$ turns out to be:
\begin{tcolorbox}[colframe=white,arc=0pt,colback=greyish2]
\vspace{-8pt}
\begin{align}
    S_{\rm anti}  &= \int dt \,  d^3 x  \bigg[ \frac{\Lambda^2}{c_s^3} \left( D_t\pi_{a}\cdot D_t\pi_{r} - c_s^2 D_i \pi_{a}\cdot D^i \pi_{r} \right) \nonumber \\
&\quad \qquad \qquad \quad  + \frac{\Lambda}{c_s^3} \, \sigma^{\mu\nu}D_{\mu}\pi_{a}\cdot \left (-\covD_{t} D_{\nu}\pi_{r}+i\left [\rho_{r},D_{\nu}\pi_{r}\right ]+\frac{i}{	\beta}D_{\nu}\pi_{a}\right ) \label{AFAction} \\
&\quad \qquad \qquad \qquad \qquad  + \frac{\sigma}{\mathcal{D}} \, \rho_{r} D_{t}\varphi_{a}  - \sigma D^{i}\varphi_{a} \left(\partial_{i}\rho_{r}+\Fcal^{r}_{ti } - \frac{i}{\beta}D_{i}\varphi_{a}\right ) \bigg] , \nonumber 
\end{align}
\vspace{-11pt}
\end{tcolorbox}
\noindent where the dot now stands for a contraction of the internal indices with a 2-dimensional Kronecker delta, $\sigma^{\mu\nu}={\rm diag}(\Sigma_{\pi}, c_s^2 \sigma_{\pi},  c_s^2 \sigma_{\pi} , c_s^2 \sigma_{\pi})$ with $\Sigma_{\pi}, \sigma_{\pi}$ both non-negative and dimensionless, and $\Lambda$ is the energy scale at which spontaneous symmetry breaking occurs.  In light of what we learned in the context of paramagnets, we have already parametrized the diffusive sector in terms of the conductivity and the diffusion coefficient. Our action doesn't contain a tadpole for the external field $A_0^a$, which is consistent with the fact that the expectation value of the $SO(3)$ Noether current must vanish for antiferromagnets. A discussion of the diffusive sector would be very similar to the analysis we carried out for paramagnets, and for this reason we'll mostly focus our attention to the $\pi$-sector in what follows. 

Based on our power counting rules, the second line of the action \eqref{AFAction} is suppressed by one power of $E/\Lambda$ compared to the first line.\footnote{We are relying on the free equation for $\pi_r$ that follows from the first line of \eqref{AFAction} to set $E \sim c_s k$.} However, we have included it because thermal effects enter first at subleading order in $E/\Lambda$. More precisely, the first two lines of our action contain terms that schematically scale as follows:
\begin{align} \label{eq: anti power counting}
	S_{\rm anti, \pi} \sim \frac{\Lambda^2}{E^2} + \frac{\Lambda}{E} \left( \frac{T}{E} + 1 \right) \ . 
\end{align}
This should be contrasted with the scaling of the terms in the leading Lagrangian for paramagnets (or the diffusive sector of antiferromagnets), where thermal effects appear already at leading order in $k/k_*$:
\begin{align}\label{eq: para power counting}
	S_{\rm para} \sim \frac{k_*}{k} \left( \frac{T}{E} + 1 \right) \ .
\end{align}
In both cases we see that at any given order in the temperature-independent expansion parameter $E/\Lambda$ or $k/k_*$ there is an additional expansion in powers of $E/T$, which we are carrying out only up to first subleading order.

Once again, varying the generating functional with respect to the external gauge fields we can calculate correlators of the currents $\mathcal{J}_r^\mu$ and $\mathcal{J}_a^\mu$.  For instance, the 2-point function of $\mathcal{J}_r^t$ along the broken directions are
\begin{align}
\langle \mathcal{J}^{tI}_{r}(\omega, \vec k)\mathcal{J}^{tJ}_{r}(-\omega, - \vec k)\rangle'&=\frac{2\Lambda}{\beta c_s}\frac{k^2 (\sigma_{\pi} \omega^{2} + \Sigma_{\pi} c_{s}^{2}k^{2})}{(\omega^{2}-c_{s}^{2}k^{2})^{2}} \, \delta^{IJ} \label{JJAntiferromagnet}
\end{align}
where $I,J = 1,2$ are here the $SO(2)$ subgroup indices, not to be confused with spatial indices. Note that, in deriving Eq. \eqref{JJAntiferromagnet} we have neglected corrections of $\mathcal{O} (E^2 /\Lambda^2)$ that cannot be trusted at the order we are working.

The retarded density-density correlator along the broken directions takes instead the form
\begin{align}
\langle \mathcal{J}^{tI}_{r}(\omega, \vec k)\mathcal{J}^{tJ}_{a}(-\omega, - \vec k)\rangle'&= \frac{\Lambda^2}{c_s} \frac{k^{2} \left[1+i(\Sigma_{\pi}- \sigma_{\pi})\omega/\Lambda\right]}{\omega^{2}-c_{s}^{2} k^{2}+ i\left(\Sigma_{\pi}\omega^{2}+\sigma_{\pi} c_s^2k^{2}\right ) \omega/\Lambda} \, \delta^{IJ}\ .
\end{align}
    The poles of this propagator are at
\begin{align}
\omega\approx \pm c_{s} k - \frac{ic_{s}^{2}k^2}{2\Lambda} (\Sigma_{\pi}+\sigma_{\pi}) \ ,
\end{align}
showing that sound modes decay at a rate $\Gamma\sim k^{2}$. The correlators above are in agreement with classic results in the literature up to the order in $E/\Lambda$ we are considering~\cite{Halperin:1969wrj, Harris:1971zza, ForsterBook}.

Our treatment of the antiferromagnet is notably different from the standard analyses, which start from the derivative expansion of the equations of motion for the total magnetization $\vec M$ and the staggered-magnetization $\vec N$ (see e.g. \cite{ForsterBook}). The effective theory \eqref{AFAction} was constructed to compute correlators of the conserved currents associated with $\vec M$ alone, and the presence of a non-trivial $\vec N$ expectation value is encoded in the assumption that $SO(3)$ is spontaneously broken down to $SO(2)$ even though the magnetization density vanishes. This last property is what distinguishes antiferromagnets from ferromagnets, which we will now turn our attention to.

\subsection{Ferromagnets\label{sec:Ferromagnets}}

Ferromagnets spontaneously break $SO(3)\to SO(2)$ because the temporal component of the Noether current acquires a non-zero expectation value, which we again assume to be along the 3-direction. As we discussed at the beginning of the previous section, what distinguishes ferromagnets from antiferromagnets is the lack of separate invariance under time reversal and discrete transformations of the form \eqref{eq: residual discrete transformations}. Only a  combination of these symmetries leaves the ground state invariant, and therefore our DKMS transformations must be amended by acting also with \eqref{eq: residual discrete transformations} on the right-hand side, thus obtaining
\begin{tcolorbox}[colframe=white,arc=0pt,colback=greyish2]
\vspace{-8pt}
\begin{subequations} \label{eq: DKMS transformations with matter fields for ferromagnets}
    \begin{align}
       & D_\mu \pi_a^1 (t) \to  - D_\mu \pi_a^1  (-t) + i \beta \nabla_t  D_\mu \pi_r^1 (-t) +  \beta \left[ \rho_r (-t) , D_\mu \pi_r^1 (-t)  \right] + \Ocal(E^2/T^2)\ , \\
       & D_\mu \pi_a^2 (t) \to  D_\mu \pi_a^2  (-t) - i \beta \nabla_t  D_\mu \pi_r^2 (-t) -  \beta \left[ \rho_r (-t) , D_\mu \pi_r^2 (-t)  \right] + \Ocal(E^2/T^2)\ , \\
       & D_\mu \pi_r^1 (t) \to  - D_\mu \pi_r^1  (-t) + \frac{i \beta}{4} \nabla_t D_\mu \pi_a^1 (-t) + \frac{\beta}{4} \left[ \rho_r (-t) , D_\mu \pi_a^1 (-t)  \right]+\Ocal(E^2/T^2)\ , \\
       & D_\mu \pi_r^2 (t) \to  D_\mu \pi_r^2  (-t) - \frac{i \beta}{4} \nabla_t D_\mu \pi_a^2 (-t) - \frac{\beta}{4} \left[ \rho_r (-t) , D_\mu \pi_a^2 (-t)  \right]+\Ocal(E^2/T^2)\ , \\
       & D_\mu \varphi_a (t) \to  - D_\mu \varphi_a  (-t) - i \beta \nabla_\mu \rho_r (-t) -i \beta \mathcal{F}_{t \mu} (-t) +  \Ocal(E^2/T^2) \ , \\
       & \rho_r (t) \to   \rho_r  (-t) + \frac{i \beta}{4} \nabla_t  D_t \varphi_a (-t) + \frac{ \beta}{4} \left[ \rho_r (-t) , D_t \varphi_a (-t)  \right] + \Ocal(E^2/T^2)\ .
    \end{align}
\end{subequations}
\vspace{-11pt}
\end{tcolorbox}
\noindent As a result, the part of the action that describes diffusion of the unbroken $SO(2)$ current now admits one additional invariant, i.e. $D_t \varphi_a$:
\begin{tcolorbox}[colframe=white,arc=0pt,colback=greyish2]
\vspace{-8pt}
\begin{align}
    S_{\rm diff}  &= \int dt \, d^3 x  \bigg[ k_\star^3 \,  D_{t}\varphi_{a} + \frac{\sigma}{\mathcal{D}} \,\rho_{r} D_{t}\varphi_{a}  - \sigma D^{i}\varphi_{a} \left(\partial_{i}\rho_{r}+\Fcal^{r}_{ti } - \frac{i}{\beta}D_{i}\varphi_{a}\right ) \bigg] . \label{eq: action ferro diffusive}
\end{align}
\vspace{-11pt}
\end{tcolorbox}
\noindent This additional term gives rise to a tadpole term for $A_0^a$, so that the expectation value of the $SO(3)$ Noether current is now no longer zero:
\begin{align} \label{eq: ferromagnet vev}
	\langle \mathcal{J}^0_A (x) \rangle = k_\star^3 \, \delta_A^3 \ .
\end{align}

This new term also gives rise to a kinetic term for the $\pi$'s with a single time derivative, since up to quadratic order in the Goldstones we have $ D_t \varphi_a = \partial_t \varphi_a + \epsilon_{IJ} \pi_a^I \partial_t \pi_r^J + \dots$. This changes the dispersion relation of the Goldstone modes, and therefore the power counting scheme. As a result, the effective action for ferromagnets is now $S_{\rm ferro}  = S_{\rm diff} + S_\pi$ with\footnote{A term of the form $\epsilon_{IJ} D_i \pi_{a}^I D^i \pi_{r}^J$ is not allowed even in ferromagnets because it is not DKMS-invariant.} 
\begin{tcolorbox}[colframe=white,arc=0pt,colback=greyish2]
\vspace{-8pt}
\begin{align}
   S_\pi &= \int dt \, d^3 x  \bigg[ - E_\star k_\star D_i \pi_{a}\cdot D^i \pi_{r} + k_\star \, \sigma_\pi D^{j}\pi_{a}\cdot \left (-\covD_{t} D_{j}\pi_{r}+i\left [\rho_{r},D_{j}\pi_{r}\right ]+\frac{i}{\beta}D_{j}\pi_{a}  \right ) \nonumber \\ 
   & \qquad \qquad \qquad \qquad \qquad \qquad \qquad \qquad \qquad \qquad \qquad \qquad \qquad \qquad + \mathcal{O}(k_\star/k) \bigg] . \label{FAction} 
\end{align}
\vspace{-11pt}
\end{tcolorbox}
This part of the action introduces a characteristic energy scale, $E_\star$, and a dimensionless coupling $\sigma_\pi$.
The leading equations of motion for $\pi_r$ now imply a quadratic dispersion relation of the form
\begin{align}
	\frac{E}{E_\star} \sim  \frac{k^2}{k_\star^2} \ ,
\end{align}
and therefore the size of the terms shown in \eqref{FAction} can be estimated to be 
\begin{align}
	S_\pi \sim \frac{k_\star^3}{k^3} + \frac{k_\star}{k}  \left( 1 + \frac{T}{E} \frac{k_\star^2}{k^2} \right)  + \mathcal{O}(k_\star/k)  \ . 
\end{align}
This scaling should be contrasted with the one for the Goldstone sector of antiferromagnets given in Eq \eqref{eq: anti power counting}. In that case, terms that appear in DKMS invariant combinations where of the same order in $E/\Lambda$, but thermal effects only appeared a subleading order in this expansion. The power counting for ferromagnets is instead more subtle, with DKMS invariant combinations containing terms of different order in $k/k_\star$, but with thermal effects now appearing at leading order. The terms of $\mathcal{O}(k_\star/k)$ that we didn't write out explicitly (e.g. $D_t \pi_r \cdot D_t \pi_a$) have been omitted to simplify our analysis, and because they would only give a subleading correction to the real part of the magnon dispersion relation. On the contrary, the terms that we included at this order because they were forced upon us by DKMS invariance provide the leading contribution to the imaginary part of the dispersion relation. 

Varying the action with respect to the external gauge fields we can first derive the Noether currents in terms of our fields, and then calculate their correlators. For brevity we are going to report here only the correlation functions of the time components of our currents:
\begin{subequations}
    \begin{align}
        \langle \mathcal{J}^{t 3}_{r}(\omega, \vec k)\mathcal{J}^{t 3}_{a}(- \omega, -\vec k)\rangle'&= \frac{\sigma k^{2}}{\omega+i\mathcal{D} k^{2}} \ , \label{eq: ferro diffusive correlator 1} \\
	   \langle \mathcal{J}^{t 3}_{r}(\omega, \vec k)\mathcal{J}^{t 3}_{r}(- \omega, -\vec k)\rangle &= k_\star^6 + \frac{2\sigma}{\beta}\frac{k^{2}}{\omega^{2}+\mathcal{D}^2 k^{4}}\ , \label{eq: ferro diffusive correlator 2}\\
      \langle \mathcal{J}^{tI}_{r}(\omega, \vec k)\mathcal{J}^{tJ}_{a}(-\omega, - \vec k)\rangle'&= \frac{i k^2 k_\star^5
   (E_\star -i  \sigma_\pi \omega) \delta^{IJ} - k_\star^7 \omega  \epsilon^{IJ}}{\omega ^2 k_\star^4 - E_\star^2 k^4 +2 i \sigma_\pi
   E_\star \omega k^4  } \ , \label{eq: ferro Goldstone correlator 1} \\
   \langle \mathcal{J}^{tI}_{r}(\omega, \vec k)\mathcal{J}^{tJ}_{r}(-\omega, - \vec k)\rangle'&= \frac{2\sigma_\pi }{\beta} \frac{ k^2 k_\star^5 \left[\delta^{IJ} \left(E_\star^2
   k^4+\omega ^2 k_\star^4\right)+2 i E_\star \omega 
   k_\star^2 k^2 \epsilon^{IJ}\right]}{  \left(\omega ^2 k_\star^4 - E_\star^2 k^4\right)^2} \ .
    \end{align} 
\end{subequations}
In the above expressions, we have only kept the terms in the numerators and denominators that can be trusted given the order in $k/k_\star$ we are working at. Note also that the Eqs. \eqref{eq: ferro diffusive correlator 1} and \eqref{eq: ferro diffusive correlator 2} are consistent with the results we found in the paramagnet section once we take into account that our Noether current now has an expectation value \eqref{eq: ferromagnet vev}. The poles in the retarded correlator \eqref{eq: ferro Goldstone correlator 1} yield once again the dispersion relations for magnon excitations, which now display a quadratic dispersion relation with a decay rate $\Gamma \sim k^4$:
\begin{align}
	\omega \simeq E_\star \left(\pm \frac{k^2}{k_\star^2} - i \sigma_\pi \frac{k^4}{k_\star^4} \right) \ .
\end{align}

\section{Conclusions\label{sec:Conclusions}}

The coset construction is heralded for its general applicability, ranging from its origins in nuclear physics \cite{Callan:1969sn} to more recent applications to systems at finite density \cite{Nicolis:2013lma,Cuomo:2020gyl}, conformal field theories~\cite{Creminelli:2014zxa, Monin:2016jmo} and gravity~\cite{Delacretaz:2014oxa, Baratella:2015yya}, just to name a few.  However, so far this technique has largely been applied to the limited case of regular effective actions for the purposes of computing scattering amplitudes or time-ordered correlators around pure states. In the present paper, we have extended the construction to Schwinger-Keldysh effective actions, which can more naturally incorporate the effects of non-trivial density matrices and facilitate the computation of a more diverse set of correlators. We focused on spontaneously broken internal symmetries, with particular emphasis on thermal states. We also  took great care in making our power counting parameters explicit, and distinguish in particular between the classical ($\hbar$) and high temperature ($E/T$) expansions. Our main conclusion is that, once the correct symmetry breaking pattern has been properly identified, the standard rules of the coset construction can be brought to bear to write down Schwinger-Keldysh effective actions. We would like to highlight in particular the advantages of the framework developed in this paper:
\begin{itemize}
    \item We retain the full non-linear structure inherent to the coset construction, thus preserving all the symmetries realized non-linearly. This should be contrasted with the common practice of linearizing Schwinger-Keldysh actions in the $a$-fields, focusing on the classical regime. Our methodology in principle allows for a systematic computation in powers of $\hbar$.
    \item In previous work on the effective Schwinger-Keldysh field theory of thermal systems, a mysterious, diffusive symmetry was needed to differentiate the normal phase from a spontaneously broken one~\cite{Glorioso:2018wxw}. In our approach, no such symmetry is needed: the symmetry breaking pattern together with basic principles such as unitarity dictate the relevant degrees of freedom and their transformation properties under all the symmetries.
\end{itemize}
In order to illustrate our framework we calculated 2-point functions of conserved spin currents   for paramagnets, anti-ferromagnets, and ferromagnets in Sec.~\ref{sec:Examples}.  Our analysis generalizes the classic work in \cite{Halperin:1969wrj, ForsterBook}, reproducing their results in the appropriate limits.

There are various avenues along which to extend the present work. First, it would be interesting to generalize our framework to include spontaneously broken space-time symmetries. This could be used to extend the coset-based approach to condensed matter systems put forward in~\cite{Nicolis:2013lma,Delacretaz:2014jka}, and would provide a different viewpoint on recent developments surrounding EFTs for dissipative hydrodynamics~\cite{Haehl:2015pja, Glorioso:2018wxw}. Second, by eschewing the classical limit, our approach could also shed a new light on the quantum properties of perfect fluids~\cite{Endlich:2010hf, Dersy:2022kjd}. Third, it would be interesting to further explore how to systematically build DKMS-invariants at higher order in $E/T$. And, finally, we would like to investigate the symmetry breaking pattern associated with finite density non-thermal density matrices, and understand how the properties of such states can be encoded in a Schwinger-Keldysh effective action. We leave all of this for future work. \\

 \noindent \textbf{Acknowledgements}: We thank Paolo Creminelli, Luca Delacretaz, Michael Landry, Mehrdad Mirbabayi, Alberto Nicolis, and Ira Rothstein for helpful discussions, and in particular Luca Delacretaz for extensive feedback on the draft. The Mathematica packages \texttt{xAct} \cite{xAct} and \texttt{xTras} \cite{Nutma:2013zea} were used extensively in the course of this work. The work of GG and RP was partially supported by the National Science Foundation under Grant No. PHY-1915611. RP is now supported by the Depatment of Energy Award DE-SC0010118. He would also like to acknowledge the hospitality of the Abdus Salam International Centre for Theoretical Physics (ICTP), where part of this work was carried out.

\appendix

\section{Schwinger-Keldysh Path Integral for a Free System at Finite Temperature\label{app:FreeInIn}}

In this Appendix, we calculate explicitly the Schwinger-Keldysh generating functional for a free, massive, relativistic $SO(N)$ scalar described by an in-out action.
\begin{align}
S = -\frac{1}{2} \int d^4 x \left( \partial_{\mu}\Phi^{A}\partial^{\mu}\Phi^{A} + m^{2} \Phi^{A}\Phi^{A} \right)\ , \qquad\quad  A  = 1,\ldots, N \ . \label{app:eq:SONScalar}
\end{align}
The goal of this appendix is twofold. First, we show explicitly how the boundary conditions give rise to off-diagonal correlation functions shown e.g. in Eq. \eqref{TwoPointFunctions}. Second, we show how the $E \ll T$ limit of the 2-point functions can be reproduced using an effective action. This is by now textbook material~\cite{KamenevBook2011}, but we discuss it here for completeness.

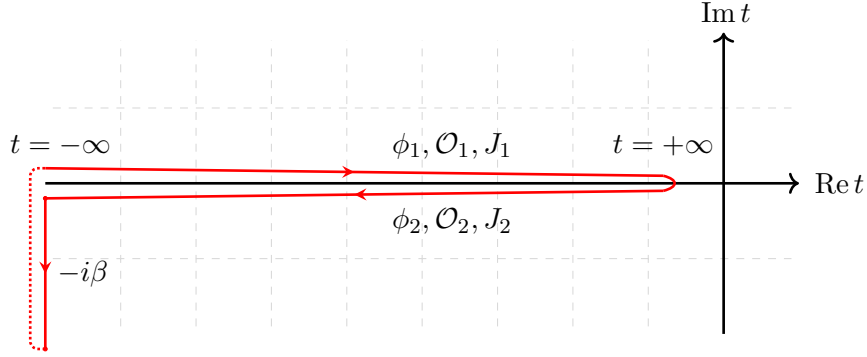
\begin{figure}[h]
 \begin{center}
 \resizebox{12cm}{!}{
\begin{tikzpicture}
\draw[help lines, color=gray!30, dashed] (-4.9,-1.9) grid (4.9,1.9);
\draw[->,line width=1pt] (-5,0)--(5,0) node[right]{$\re t$};
\draw[->,line width=1pt] (4,-2)--(4,2) node[above]{$\im t$};
\coordinate (TL) at (-5,.2);
\coordinate (BL) at (-5,-.2);
\coordinate (TR) at (3.2,.1);
\coordinate (BR) at (3.2,-.1);
\coordinate (BBL) at (-5,-2.2);
\begin{scope}[decoration={
    markings,
    mark=at position 0.5 with {\arrow{stealth}}}
    ] 
    \draw[postaction={decorate},line width=1pt, red] (BL)--(BBL){};
    \draw[postaction={decorate},line width=1pt, red] (BR)--(BL){};
	\draw[postaction={decorate},line width=1pt, red] (TL)--(TR){};
	\draw[line width=1pt, red] (BR).. controls ($(BR)+(.2,.05)$) and ($(TR)+(.2,-.05)$)..(TR){};
	\draw[line width=1pt, red,densely dotted,rounded corners=3] (BBL)--($(BBL)+(-.2,0)$)--($(TL)+(-.2,0)$)--(TL){};
\end{scope}

\node at (3.2,.5) [minimum size=.2cm] {$t=+\infty$};
\node at (-4.8,.5) [minimum size=.2cm] {$t=-\infty$};
\node at (.4,-.5) [minimum size=.2cm] {$\phi_{2},\Ocal_{2},J_{2}$};
\node at (.4,.5) [minimum size=.2cm] {$\phi_{1},\Ocal_{1},J_{1}$};

\node at (-4.5,-1.2) [minimum size=.2cm] {$-i\beta$};
\filldraw[color=red] (-5,-.2) circle (.025);
\filldraw[color=red] (-5,-2.2) circle (.025);
\end{tikzpicture}
}
\caption{The contour $\mathcal{C}$ used for the Schwinger-Keldysh generator \eqref{ConnectedGeneratorPIInsertions} when $\rho\propto e^{-\beta H}$. The dotted line represents the completion of the trace; there is no evolution along this segment.  Other contours for finite-$T$ systems can be found in the literature; see \cite{Kraemmer:2003gd} for a review.}\label{fig:InInTimeContourFiniteBeta}
\end{center}
 \end{figure}

Our starting point is the generating functional in Eq. \eqref{ConnectedGeneratorPIInsertions} with $\varphi = \Ocal= \vec \Phi$:

\begin{align} \label{eq: SO(N) generating functional}
Z[J_{1},J_{2}] &= \! \int \! D\vec{\Phi}_{a} D \vec{\Phi}_{b} D \vec{\Phi}_{c}\,\bra{\vec{\Phi}_{a}, -\infty} \rho\ket{\vec{\Phi}_{c}, -\infty}\bra{\vec{\Phi}_{c}, -\infty}\bar{T} e^{-i \int \vec J_{2}\cdot \vec \Phi}\ket{\vec{\Phi}_{b}, +\infty} \\ 
& \qquad \qquad \qquad \qquad \qquad \qquad \qquad \qquad \qquad \times \bra{\vec{\Phi}_{b}, +\infty}T e^{i \int \vec J_{1} \cdot \vec \Phi}\ket{\vec{\Phi}_{a}, -\infty} , \nonumber
\end{align}
When the density matrix is thermal, $\rho\propto e^{-\beta H}$, this functional can be computed from the knowledge of the amplitude
\begin{align}
\bra{\vec{\Phi}_{f}(\bfx), + \tts}  T e^{i \int\rd^4 x\, \vec{J}\cdot\vec{\Phi}}\ket{\vec{\Phi}_{i}(\bfx), -\tts}&\equiv \int_{\vec{\Phi}(-\tts)=\vec{\Phi}_{i}(\bfx)}^{\vec{\Phi}(+\tts)=\vec{\Phi}_{f}(\bfx)}\Dcal\Phi\, e^{iS[\vec{\Phi}]+i\int\rd^4x\,\vec{J}\cdot\vec{\Phi}}\nn
& \equiv \langle \vec{\Phi}_{f}, + \tts|\vec{\Phi}_{i}, - \tts\rangle_{\vec{J}}\ ,\label{app:eq:SONAmplitude}
\end{align}
since each factor on the right-hand side of \eqref{eq: SO(N) generating functional} is a special case of the quantity above.  In particular, the thermal density matrix factor comes from setting the source $\vec{J}$ to zero and properly Wick rotating. Combining all three factors together, we obtain a single path integral expression for the generating functional, with fields defined along a time-contour $\mathcal{C}$ in the complex plane shown in Fig.~\ref{fig:InInTimeContourFiniteBeta}.

Due to the free nature of \eqref{app:eq:SONScalar}, the amplitude in \eqref{app:eq:SONAmplitude} can be computed exactly by substituting the classical solution obeying the appropriate boundary conditions in the presence of the source $\vec{J}$ into the action.  The explicit form of such a classical solution is\footnote{Note that in this appendix we are denoting spatial vectors in boldface to reserve the vector symbol for objects in the fundamental representation of the internal $SO(N)$ group.}
\begin{subequations}
\begin{align}
& \vec{\Phi}_{\rm cl}(t,\bfk )=K(t,\bfk,\tts )\cdot \vec{\Phi}_{b}(\bfk )+K(-t,\bfk,\tts )\cdot \vec{\Phi}_{a}(\bfk )+\int_{-\tts }^{\tts }\rd t'\, G(t,t',\bfk,\tts)\cdot\vec{J}(t',\bfk ) \ , \\
& K^{AB}(t,\bfk,\tts ) \equiv \delta^{AB}\frac{\sin(\omega_{\bfk }(t+\tts ))}{\sin(2\omega_{\bfk }\tts )} \ , \\
& G^{AB}(t,t',\bfk,\tts)\equiv\delta^{AB}\times\Big[ \frac{\sin(\omega_{\bfk }(t-t'))}{2\omega_{\bfk }}\left [\theta(t-t')-\theta(t'-t)\right ] \nonumber \\
& \qquad \qquad \qquad \qquad \quad-\frac{2\cos(\omega_{\bfk}(t+t'))-\cos(\omega_{\bfk}(2\tts +t-t'))-\cos(\omega_{\bfk}(2\tts +t'-t))}{4\omega_{\bfk }\sin(2\omega_{\bfk }\tts )}\Big]\ ,\label{app:eq:FreeThermalPropagators}
\end{align}
\end{subequations}
where $\omega_{\bfk}^{2}=\bfk^{2}+m^{2}$ and where, for now, we are imposing boundary conditions at $\pm \tts $. The $G^{AB}(t,t',\bfk,\tts)$ propagator obeys $(\square_x -m^2)G^{AB}(x,z,\tts)=-\delta^{AB}\delta^4(x-z)$ in position-space and is symmetric under $t\to t'$ and vanishes at the boundaries where either $t'$ or $t$ equals $\pm \tts $ while $K^{AB}(x, \bfz,\tts )$ solves the equation of motion identically, reduces to the identity at the $t=\tts $ boundary, and vanishes at $t=-\tts $.  

Plugging the expression above into the action, the amplitude \eqref{app:eq:SONAmplitude} reduces to
\begin{align}
\langle \vec{\Phi}_{f}, + \tts|\vec{\Phi}_{i}, - \tts\rangle_{\vec{J}}&=\exp\Big[\frac{i}{2}\int \frac{\rd^3k}{(2 \pi)^3} \, \omega_{\bfk}\cot(2\omega_{\bfk} \tts)\left (\vec{\Phi}_{i}\cdot \vec{\Phi}_{i}+\vec{\Phi}_{f}\cdot \vec{\Phi}_{f}\right )-2 \omega_{\bfk}\csc(2\omega_{\bfk} \tts)\vec{\Phi}_{i}\cdot \vec{\Phi}_{f}\nn
&\quad + i\int_{-\tts}^{\tts}\rd t\, \int  \frac{\rd^3k}{(2 \pi)^3} \, \vec{J}(t,\bfk)\cdot K(t,\bfk,\tts)\cdot \vec{\Phi}_{f}+\vec{J}(t,\bfk)\cdot K(-t,\bfk,\tts)\cdot \vec{\Phi}_{i}\nn
&\quad  +\frac{i}{2}\int_{-\tts}^{\tts}\rd t\rd w \int \frac{\rd^3k}{(2 \pi)^3} \, \vec{J}(t,-\bfk)\cdot G(t,w,\bfk)\cdot \vec{J}(w,\bfk)\Big]\label{app:eq:ExplicitSONAmplitude}
\end{align}
where $\vec{\Phi}_{i}=\vec{\Phi}_{i}(\bfk)$ everywhere. Note that this result is manifestly invariant under $SO(N)$. The density matrix components come from setting $\tts=i\beta/2$ and $\vec{J}\to 0$ in the above.

After calculating the three amplitude factors in \eqref{eq: SO(N) generating functional}, it is straightforward to stitch them all together by computing the remaining Gaussian path integrals over $\vec \Phi_a, \vec \Phi_b$, and $\vec \Phi_c$. It is this last step that gives rise to the cross terms $\sim J_1\times J_2$. The ultimate expression is written most compactly in terms of the $\vec{J}_{a}, \vec{J}_{r}$ Keldysh basis sources:
\begin{align}
&\quad \ln Z[\vec{J}_{a},\vec{J}_{r}]\nn
&=-\frac{1}{2}\int\frac{\rd\omega\rd^3k}{(2 \pi)^4} \,  \begin{pmatrix}
\vec{J}_{r}(-k) & \vec{J}_{a}(-k)
\end{pmatrix}
\cdot
\begin{pmatrix}
0& \frac{i}{(\omega+i\epsilon)^{2}-\omega_{\bfk}^{2}}\\
 \frac{i}{(\omega-i\epsilon)^{2}-\omega_{\bfk}^{2}} & \left (\frac{1}{2}+\frac{1}{e^{\beta|\omega|}-1}\right ) 2 \pi\delta (\omega^{2}-\omega_{\bfk}^{2}) 
\end{pmatrix}
\cdot \begin{pmatrix}
\vec{J}_{r}(k) \\ \vec{J}_{a}(k)
\end{pmatrix}\ .\label{app:eq:FreeSKGaussianResultFourierFiniteT}
\end{align}

In this paper we considered an alternative representation of the generating functional that relies on \textit{effective} fields\footnote{In what follows, we  denote these effective fields with the same symbol, $\vec \Phi$, to streamline the notation.} and for which the usual, naive rules of Gaussian integration can be used.  Such a representation allows us to avoid the complicated, multi-step process above.  From this viewpoint, the generator  \eqref{app:eq:FreeSKGaussianResultFourierFiniteT} is instead constructed as
 \begin{align}
  Z[J_{a},J_{r}] &=\int\Dcal\vec{ \Phi}_{a}\Dcal\vec{ \Phi}_{r}\, e^{i\SEFT[\vec{ \Phi}_{a},\vec{ \Phi}_{r}]+i\int\rd^{d+1}x\, \vec{J}_{r}\cdot\vec{ \Phi}_{a}+\vec{J}_{a}\cdot\vec{ \Phi}_{r}}
 \end{align}
 for some $\SEFT$.  Finding the appropriate effective action which reproduces \eqref{app:eq:FreeSKGaussianResultFourierFiniteT} is a simple exercise in reverse engineering via standard Gaussian integral formulas, and a convenient form is\footnote{Equivalence follows from the identity $\lim_{\varepsilon\to 0}\frac{\varepsilon}{x^{2}+\varepsilon^{2}}=\pi\delta(x)$.  The fact that the imaginary terms are $\Ocal(\varepsilon)$ is an artifact of the free limit \cite{KamenevBook2011}; they are finite in realistic, interacting systems, see e.g.~\cite{Grozdanov:2013dba}.}
 \begin{align}
 \SEFT[\vec{ \Phi}_{a},\vec{ \Phi}_{r}]&\equiv \int\rd^4x\, \left (-\partial\vec{ \Phi}_{a}\cdot\partial\vec{ \Phi}_{r}-m^{2}\vec{ \Phi}_{a}\cdot\vec{ \Phi}_{r}- 2\varepsilon\vec{\Phi}_{a}(x)\cdot\partial_{t}\vec{ \Phi}_{r}(x) \right ) \nn
 &\quad \qquad \qquad \qquad + i\varepsilon  \int\frac{\rd\omega\rd^3k}{(2 \pi)^4} \, \omega\coth(\beta\omega/2)\vec{ \Phi}_{a}(k) \cdot\vec{ \Phi}_{a}(-k) \ ,\label{app:eq:FreeEffectiveAction}
 \end{align}
 where we take $\varepsilon\to 0^{+}$ at the end of any given computation, as usual.

For emphasis, in contrast to the path integral considered at the start of this section, when using \eqref{app:eq:FreeEffectiveAction} there is no need to carefully consider boundary conditions on fields at $t\to \pm \infty$ or the presence of a non-trivial density matrix.  Such features are already encoded in $\SEFT$ itself. In particular, the explicit factors of $\beta$ in \eqref{app:eq:FreeEffectiveAction} reflect the thermal nature of the system which is also a consequence of the dynamical KMS symmetry \cite{Glorioso:2018wxw} of the effective $\vec{\Phi}$ fields.  Explicitly, this acts linearly on the fields in frequency space as
\begin{subequations} \label{app:eq:DKMSFree}
\begin{align}
   \vec{\Phi}_{a}(\omega)&\to  \cosh (\beta \omega / 2) \vec{\Phi}_{a}(-\omega) - 2  \sinh (\beta \omega / 2) \vec{\Phi}_{r}(-\omega) \\
   \vec{\Phi}_{r}(\omega)&\to\cosh (\beta \omega / 2) \vec{\Phi}_{r} (-\omega) - \frac{1}{2} \sinh (\beta \omega / 2)  \vec{\Phi}_{a}(-\omega)	\ ,
\end{align}
\end{subequations}
 and it can be checked that \eqref{app:eq:FreeEffectiveAction} is precisely invariant under the above.
   
For general values of $\beta\omega$, the representation of the generating functional via an effective action \eqref{app:eq:FreeEffectiveAction} is not obviously advantageous.  Though this construction allows us to more easily use familiar path-integral methods, its convenience is offset by the fact that the  terms $\sim \vec{\Phi}_{a}^{2}$ are non-local.   However, $\SEFT$ becomes approximately local in the low-energy limit, $\beta\omega\ll 1$:
\begin{align}
       \SEFT& \simeq\int\rd^4x\, \left\{ -\partial\vec{\Phi}_{a}\cdot\partial\vec{\Phi}_{r}-m^{2}\vec{\Phi}_{a}\cdot\vec{\Phi}_{r}+ 2\varepsilon\left( -\vec{\Phi}_{a}\cdot\partial_{t}\vec{\Phi}_{r}+\frac{i}{\beta}\vec{\Phi}_{a}(x)\cdot\vec{\Phi}_{a}(x)\right) \right\} \ .\label{app:eq:FreeEffectiveActionDissipativeTerm}
   \end{align}
   The structure of the $\Ocal(\varepsilon)$ terms is dictated by the $\beta$-expansion of the dynamical KMS symmetry \eqref{app:eq:DKMSFree} in which the symmetry acts on the fields in momentum space as in
   \begin{align}
   \vec{\Phi}_{a}(\omega,\bfk)\to \vec{\Phi}_{a}(-\omega,\bfk)-\beta\omega \vec{\Phi}_{r}(-\omega,\bfk) \ , \qquad \vec{\Phi}_{r}(\omega,\bfk)\to  \vec{\Phi}_{r}(-\omega,\bfk)  - \frac{\beta \omega}{4} \vec{\Phi}_{a}(-\omega,\bfk)\ ,\label{app:eq:DKMSFreeHbarExpansion}
   \end{align}
   under which the $\Ocal(\varepsilon)$ terms in \eqref{app:eq:FreeEffectiveActionDissipativeTerm} are strictly invariant.  
   
Note that the effective action is invariant under two copies of the global $SO(N)$ symmetries provided we disregard the terms proportional to $\varepsilon$, as is customary when discussing the symmetries of effective actions. However, it is easy to show that both copies of the symmetries are realized linearly on our effective fields, meaning that the off-diagonal symmetry is not spontaneously broken in this very simple case. This is an artifact of the free limit, which prevents the existence of a hydrodynamic regime at low energies.

\section{Correlator Cheat Sheet\label{app:CorrelatorCheatSheet}}

A wide variety of conventions and notations exist for the possible correlators in quantum field theory. Ours are found below, along with  various useful relations they satisfy.  Given a set of operators $\Ocal^{i}$ with $i$ some general indices, some of the correlators one might be interested in calculating are 
\begin{align}
\Delta^{ij}(x,x')&\equiv \langle \left [\Ocal^{i}(x),\Ocal^{j}(x')\right ]\rangle \nn
G^{ij}_{W}(x,x')&\equiv \langle \Ocal^{i}(x)\Ocal^{j}(x')\rangle\nn
G^{ij}_{R}(x,x')&\equiv i\theta(t-t')\langle \left [\Ocal^{i}(x),\Ocal^{j}(x')\right ]\rangle = i \langle \Ocal_{r}^{i}(x)\Ocal_{a}^{j}(x')\rangle\nn
G^{ij}_{A}(x,x')&\equiv -i\theta(t'-t)\langle \left [\Ocal^{i}(x),\Ocal^{j}(x')\right ]\rangle=-i\langle \Ocal_{a}^{i}(x)\Ocal_{r}^{j}(x')\rangle\nn
G^{ij}_{S}(x,x')&\equiv \frac{1}{2}\langle \{\Ocal^{i}(x),\Ocal^{j}(x')\}\rangle= \langle \Ocal_{r}^{i}(x)\Ocal_{r}^{j}(x')\rangle\ ,
\end{align}
where for the last three correlators we have also provided an expression in terms of fields in the Keldysh basis introduced in Sec.~\ref{sec:KeldyshRotation}.  The momentum-space versions of the above are defined in the usual way, with $G^{ij}_{X}(\omega,\vec k)$ determined from a correlator of $\Ocal^{i}(\omega,\vec k)$ and $\Ocal^{j}(-\omega,-\vec k)$.

The commutator $\Delta^{ij}$ and the retarded correlator $G_{R}^{ij}$ are related in momentum space by 
\begin{align}
G_{R}^{ij}(\omega,\vec k)&=\lim _{\varepsilon\to 0^{+}}\int \frac{\rd\omega'}{2 \pi}\, \frac{\Delta^{ij}(\omega',\vec k)}{\omega'-\omega-i\varepsilon}\ ,\label{app:eq:RetardedFromCommutator}
\end{align}
as follows from the Fourier-representation of the Heaviside function. The zero-frequency limit of this relation determines the \textit{static susceptibilities} $\chi^{ij}(\vec k)$,
   \begin{align}
 \chi^{ij}(\vec k)\equiv \lim_{\omega\to 0}G_{R}^{ij}(\omega,\vec k\neq 0)= \lim _{\varepsilon\to 0^{+}}\int \frac{\rd\omega'}{2 \pi}\, \frac{\Delta^{ij}(\omega',\vec k)}{\omega'-i\varepsilon}\ .\label{app:eq:StaticSusceptibilities}
\end{align}

At finite temperature, the KMS conditions relate various correlators to each other.  These can be straightforwardly derived starting from inserting ${\bf 1}=e^{-\beta H}e^{\beta H}$ judiciously into the Wightman function $G_{W}^{ij}$ (also called the \textit{dynamical structure factor}) and using the cyclicity of the trace to get (highlighting only the temporal and frequency dependence)
\begin{align}
G^{ij}_{W}(t)&=G^{ji}_{W}(-i\beta-t)= e^{i\beta\partial_{t}}G^{ji}_{W}(-t) \qquad 
\implies \qquad G^{ij}_{W}(\omega) =e^{\beta\omega}G^{ji}_{W}(-\omega)\ .
\end{align}
In particular, in the limit $\beta\omega\ll 1$ that is the focus of our paper, this implies a form of the Fluctuation-Dissipation Theorem
\begin{align}
G_{S}^{ij}(\omega)\simeq G_{W}^{ij}(\omega)\approx \frac{\Delta^{ij}(\omega)}{\beta\omega} \ ,\label{app:eq:ApproximateKMSRelations}
\end{align}
up to corrections of higher order in $\beta\omega$. The first relation above implies that $\Ocal^{i}$ and $\Ocal^{j}$ approximately commute, thereby motivating the ``classical limit" terminology.   

Using the effective actions discussed in this paper it is straightforward to calculate the symmetric $G_{S}^{ij}$ viewed as an $r$-$r$ correlator in the language of the Keldysh rotation of Sec.~\ref{sec:KeldyshRotation}. The low energy expressions for $G_{R}^{ij}, \Delta^{ij}$, and $G_{W}^{ij}$ then follow from $G_{S}^{ij}$using \eqref{app:eq:RetardedFromCommutator} and \eqref{app:eq:ApproximateKMSRelations}.

\bibliographystyle{utphys}
\bibliography{Bibliography}

\end{document}